\documentstyle[aaspp4,tighten,flushrt]{article}
\received{}
\accepted{}

\lefthead{Basu, Johnstone, Martin}
\righthead{Evolution of an Expanding Superbubble}

\begin{document}

\title{Dynamical Evolution and Ionization Structure of an
Expanding Superbubble: Application to W4} 

\vspace{0.2in}
\author{Shantanu Basu, Doug Johnstone, and P. G. Martin}
\affil{Canadian Institute for Theoretical Astrophysics,
University of Toronto, 60 St. George Street, Toronto, Ontario M5S 3H8, Canada;
basu, johnstone, pgmartin@cita.utoronto.ca}
\authoremail{basu@cita.utoronto.ca}

\vspace{0.3in}
\centerline{to appear in {\it The Astrophysical Journal}}
\vspace{0.1in}

\newcommand{\HI}{\ion{H}{1}~}
\newcommand{\HII}{\ion{H}{2}~}
\newcommand{\Rs}{R_{\rm s}}
\newcommand{\Rst}{R_{\rm St}}
\newcommand{\yt}{\tilde{y}}
\newcommand{\ttil}{\tilde{t}}
\newcommand{\tN}{\tilde{N}}
\newcommand{\ts}{\tilde{s}}
\newcommand{\tP}{\tilde{P}}
\newcommand{\bl}[1]{\mbox{\boldmath$ #1 $}}
\newcommand{\beq}{\begin{equation}}
\newcommand{\eeq}{\end{equation}}
\renewcommand{\and}{\&\ }
\newcommand{\refer}{\reference}

\newcommand{\halpha}{\mbox{H$\alpha$}~}

\begin{abstract}

Recent observations have revealed a superbubble associated
with the young stellar cluster OCl 352 near the W4 \HII region:
a void in \HI emission (Normandeau, Taylor, \& Dewdney), and a
bright shell in \halpha emission (Dennison, Topasna, \&
Simonetti). We investigate
the hypothesis that the bubble is blown by stellar winds from
the O-type stars in the association. The Kompaneets approximation
is adapted to model a wind-blown bubble in a stratified interstellar medium.
We describe some general principles necessary for understanding
the dynamics of an expanding bubble and the associated ionization structure
in a stratified atmosphere. The Kompaneets model can be used to
determine the mean scale height of the ambient
medium as well as the age of the bubble. The ionization
structure also places constraints on the ambient density near the cluster.
We also estimate the surface brightness of the shell and the
fraction of ionizing photons which escape the bubble.
The prescription we use can be applied to any observed bubble that is
blown by the effectively continuous energy output of stellar winds or
multiple supernovae. Application to the W4 superbubble shows that the
mean scale height of the ambient gas around
the cluster is remarkably small, 25 pc for a cluster distance of 
2.35 kpc. The age
of the bubble is estimated to be about 2.5 Myr, consistent with
the notion that the bubble is blown by stellar winds from a very young
cluster in which no supernovae have yet occurred.

\end{abstract}


\section{Introduction}

Normandeau, Taylor, \& Dewdney (1996, 1997; hereafter NTD) 
have observed a large cavity in \HI above the W4 \HII region in
the Pilot project of the arcminute resolution Canadian Galactic Plane
Survey (CGPS). The conical cavity, found in the velocity range
$v_{\rm LSR} =  -31.9$ to $-46.7$ km s$^{-1}$, opens upward
above (and perpendicular to) the Galactic plane, and is bounded below 
by the W4 \HII region, which is part of the larger W3/W4/W5 \HII region complex 
located in the Perseus arm of the outer Galaxy. 
Panoramic images of CO and thermal dust (IRAS) emission from this region,
also at arcminute resolution, have been presented by Heyer \& Terebey (1998).
An association of nine O-type stars, part of the stellar cluster OCl 352 
(also IC 1805), centered at ($l,b$) = ($134.7\arcdeg,0.9\arcdeg$), 
is situated near the base of the cavity, suggesting that the cavity has 
been created by outflows from these stars.
Spectroscopic parallax (Massey, Johnson, \& DeGioia-Eastwood 1995)
places OCl 352 at a distance 2.35 kpc; at this distance, the \HI cavity 
has a width $\simeq$ 110 pc at a latitude $b \simeq 3.5\arcdeg$, or 
$\simeq 110$ pc above the cluster.
Radio continuum data presented by NTD show the presence of a
shell of ionized gas along the lower edge of the \HI cavity.
Dennison, Topasna, \& Simonetti (1997; hereafter DTS) 
have observed the same region in H$\alpha$
with a field of view extending to higher Galactic latitude. Their image
shows that the \HI cavity of NTD is located near the base of a highly elongated
shell of H$\alpha$ emission which appears to close far above the
Galactic plane.
This shell is interpreted here to be the dense ionized wall
of swept-up gas surrounding a bubble of hot rarefied gas created by
the outflow from the massive stars.

This structure is one example of a vast collection of bubbles and
associated shells that are thought to populate the interstellar medium
(ISM) of galaxies. In our Galaxy, \HI observations (Heiles 1979, 1984)
reveal numerous shells with dimensions of several hundred parsecs.
Other spiral, irregular, and dwarf galaxies also show evidence for
shells and holes in \HI (e.g., Brinks \& Bajaja 1986; Deul \& den Hartog 1990;
Puche et al. 1992). The holes in \HI may be sites where the bubbles have
blown out of the stratified ISM and into the halo of the host galaxies.
Norman \& Ikeuchi (1989) suggest that these ``chimneys'' are a common
feature in our Galaxy.

The well-resolved structure of the W4 superbubble and the well known
properties of the stellar cluster OCl 352 
that drives the bubble afford us a chance to apply theoretical models
to gain insight into the massive star-ISM interaction in a star-forming
region of our Galaxy. We show that the highly elongated nature of the 
superbubble implies the presence of significant vertical stratification in 
the ambient ISM; the scale height is remarkably small.

Theoretical models of bubble expansion in a stratified medium come in 
many varieties. Kompaneets (1960) found a semianalytic solution for the
expansion of a blast wave in an exponential atmosphere. 
The Kompaneets approximation assumes the following: 1)
uniform pressure within the bubble, 2) bubble expansion in a
direction normal to the local surface, and 3) expansion speed 
implied by a strong shock (i.e., the internal pressure dominates
the external pressure). Kompaneets found an analytic expression for the
shape of the bubble during its expansion. The time-evolution of the 
bubble can be found by numerical integration.
The Kompaneets solution has been used and adapted for application to various 
astrophysical phenomena, including relativistic blast waves (Shapiro 1979),
active galaxy winds (Schiano 1985), and impacts within the deep gaseous
envelopes of giant planets (Korycansky 1992). 
More sophisticated models of bubble expansion include the thin-shell
approximation (MacLow \& McCray 1988; Bisnovatyi-Kogan, Blinnikov, \&
Silich 1989), which employs the first two assumptions of the Kompaneets
approximation, but determines the expansion speed through direct
numerical integration of the momentum equation 
for various segments of the thin shell of swept-up gas.
This numerical approach accounts for the inertia of the swept-up shell, 
and external pressure and gravity can also be included.
Finally, full numerical integration of the hydrodynamic equations
(e.g., Tomisaka \& Ikeuchi 1986; MacLow, McCray, \& Norman 1989;
Tenorio-Tagle, Rozyczka, \& Bodenheimer 1990) and 
magnetohydrodynamic equations (Tomisaka 1992, 1998) yield
the most complete solutions to date. For a review of
models of bubble expansion in the interstellar medium, see 
Bisnovatyi-Kogan \& Silich (1995).

In this paper, we choose to apply the Kompaneets model because,
in addition to its simplicity, 
the semianalytic solution allows a 
more direct insight into the physics of bubble expansion. An observed
bubble can be matched to theory using analytic expressions, yielding
straightforward estimates of the atmospheric structure and bubble age.
The original Kompaneets model is adapted so that the energy input is
continuous rather than occurring at an initial instant, as is expected
with stellar winds or multiple sequential supernovae.
We discuss the conditions under which the Kompaneets solution is
valid, and show how some of its properties can be
used in an analysis of observed bubbles.
Application of the model to the W4 superbubble gives us a more accurate age
estimate than is possible with models of spherical expansion, as well as a
dynamical estimate of the gas scale height in the W4 star-forming region.
Given the structure of the bubble and ambient atmosphere,
we also illustrate the evolution of the ionized gas structure
created by the O-type stars. The current structure of the ionized
shell places an additional constraint on the mean density of the 
ambient atmosphere and allows predictions of how much ionizing flux
escapes through the bubble.

The outline of this paper is as follows. In \S\ 2 we apply
the Kompaneets model to the structure and
evolution of the wind-blown bubble. The complementary
role of the ionizing UV flux from the massive stars is discussed
in \S\ 3.  We interpret our results in \S\ 4, and a summary of our main
conclusions is given in \S\ 5.
Readers who wish to review a more complete exposition of the Kompaneets 
solution are referred to the Appendix.

\section{Dynamical Evolution of a Superbubble}

\subsection{Early Evolution and Blowout Condition}

\label{sec_earlyevol}

The early evolution of a superbubble in a stratified medium 
can be described by the model for spherical expansion of a stellar-wind
bubble in a homogeneous medium, as developed by Castor, McCray, \& 
Weaver (1975) and Weaver et al. (1977).
Their model can be applied to bubbles powered by the combined 
winds of several O and B stars if they are closely associated, and
can even apply to bubbles powered by multiple sequential supernovae
from such an association, since the energy output is effectively
continuous (McCray \& Kafatos 1987; MacLow \& McCray 1988).
We briefly describe the model of Castor et al. (1975) and 
Weaver et al. (1977) before proceeding to discuss the effects of 
vertical stratification.

The earliest phase is one of free expansion of the wind, which drives
a shock front into the ambient ISM. A shell of swept-up ISM is formed
behind the shock front, and once its inertia 
slows down the shell appreciably, an inner shock front is also formed.
This inner shock slows down the hypersonic free wind before it encounters
the more slowly moving shell, creating a region of hot, shocked wind gas.
Hence, most of the bubble evolution is characterized by a four-zone
structure: (a) an innermost freely expanding stellar wind surrounded by
the inner shock surface, (b) a hot, almost isobaric region consisting
of shocked wind gas and a small fraction of swept-up ISM gas, (c)
a shell of swept-up ISM gas separated from region (b) by a
contact discontinuity, and (d) the undisturbed ISM, separated from 
region (c) by the outer shock front.
Radiative cooling causes region (c) to collapse into a thin, dense shell
within a few thousand years, while cooling remains relatively 
unimportant in the rarefied region (b). The inner shock front radius
$R_1$ is found to be much smaller than the outer shock front radius
$R_2$. Hence, the expansion of the bubble can be modeled effectively as
that of a thin-shell driven into the ambient medium with
uniform density $\rho_0$ by an isobaric bubble of radius $\Rs = R_2$. 
The internal energy of the bubble $E_{\rm th}$ (and hence the mean 
pressure $P$) is determined
by the energy input from the luminosity $L_0$ of the wind minus the
work done in expanding the bubble.
Under these conditions, Castor et al. (1975) find the analytic solution
\begin{eqnarray}
\Rs & = & \left( \frac{125}{154 \pi} \right)^{1/5} \, L_0^{1/5} \, \rho_0^{-1/5}
\, t^{3/5}, \label{Rs}\\
E_{\rm th} & = & \frac{5}{11}L_0 t, \label{Eth}\\
P & = & \frac{2}{3} \frac{E_{\rm th}}{\Omega} = \frac{7}{(3850 \pi)^{2/5}} \,
L_0^{2/5} \, \rho_0^{3/5} \, t^{-4/5}, \label{Press}
\end{eqnarray}
where $\Omega = 4 \pi \Rs^3/3$ is the volume of the bubble, and 
a ratio of specific heats $\gamma = 5/3$ has been used for the equation 
of state in the bubble interior.
The expansion will dissipate and the shell will begin to merge with
the interstellar medium when the expansion speed $u = \frac{3}{5} \Rs/t$ 
($\propto t^{-2/5}$) decreases to the sound speed of the external medium
$c_{\rm s,e} = (P_{\rm e}/\rho_0)^{1/2}$, where $P_{\rm e}$ is the external 
pressure (essentially the same time that $P=P_{\rm e}$).
This occurs at the stalling radius
\beq
\label{Rstall}
R_{\rm stall} = \left( \frac{27}{154 \pi} \right)^{1/2} 
L_0^{1/2} \rho_0^{1/4} P_{\rm e}^{-3/4}.
\eeq

In a stratified medium, $R_{\rm stall}$ is crucial in determining
whether a bubble will ``blow out''. In such a 
medium, the initial expansion will be spherical, as in a homogeneous
atmosphere. However, if the expansion does not stall before proceeding 
beyond one scale height $H$ of an exponential atmosphere, then it will sense
the stratification and the top of the shell will begin to accelerate and
reach an infinite height in a finite time, which defines blowout.
Hence, the dimensionless ratio $b = R_{\rm stall}/H$ is an important diagnostic
for bubbles in a stratified medium; we surmise that $b \gtrsim 1$ for blowout 
to occur. Indeed, MacLow \& McCray (1988) find a similar dimensionless ratio 
$D$ which gauges whether or not their bubbles blow out of a finite pressure
exponential medium. Their parameter $D$ can be written in our terminology
as
\beq
D = \left( \frac{154 \pi}{27}\right) b^2 = 17.9 \; b^2.
\eeq
MacLow \& McCray (1988) find that a model with $D=10$ ($b=0.75$) is confined 
but one with $D=100$ ($b=2.4$) blows out of the atmosphere, consistent with 
our expectation that blowout occurs for $b \gtrsim 1$.

In the case of the W4 superbubble, which can be blown by stellar
winds alone, the estimated wind luminosity of the driving cluster
OCl 352 is $L_0 = 3 \times 10^{37}$ ergs s$^{-1}$ (NTD).
Hence, for this bubble, equation (\ref{Rstall}) gives
\beq
\label{blowout}
b = 4 \, \left( \frac{L_0}{3 \times 10^{37} \, 
{\rm ergs \, s}^{-1}} \right)^{1/2} \:
\left( \frac{n_0}{1 \, {\rm cm}^{-3}} \right)^{1/4} \:
\left( \frac{10^4  \, {\rm cm}^{-3}\, {\rm K}}{P_{\rm e}/k} \right)^{3/4} \:
\left( \frac{100 \, {\rm pc}}{H} \right),
\eeq
so that any standard interstellar values for $n_0, P_{\rm e}$, and $H$
will yield $b > 1$. In the W4 region, probably $b \gg 1$, since 
the scale height $H$ is actually much less than 100 pc (see \S\ 2), and
the ambient density $n_0$ is actually greater than 1 cm$^{-3}$ (see \S\ 3).
Our adopted normalizing value of $P_{\rm e}/k$ also represents the high 
end of estimated pressures in the ISM (Kulkarni \& Heiles 1987).

\subsection{Expansion in a Stratified Atmosphere: Application to W4}

Kompaneets (1960) developed a semianalytic solution for shock wave 
propagation in an exponential atmosphere. This solution was originally
used to study blast wave propagation, but we adapt it to include a 
continuous input of energy, as also done by Schiano (1985) in his
study of active galaxy winds. The model gives a good idea of how shock
waves propagate in a stratified medium without having to solve the
full hydrodynamic equations numerically. The shape of the bubble is 
determined analytically, and the time evolution is determined by 
integrating a differential equation. The spatial and temporal solutions
of the Komapneets model for a wind-blown bubble are presented in 
Appendix A. In the following sections, we apply the solution to
the W4 superbubble.

\subsubsection{Relation of the Spatial Profiles to the W4 Superbubble}

Figure 1 shows the surface of the shock front $r(z,y)$ (eq. [\ref{kompsoln}])
for various values of the dimensionless quantity $\yt = y/H$, which
parametrizes the evolution of the bubble (see \S\ \ref{apdx_spatial}). 
The shock surface is spherical at early times,
but becomes increasingly elongated in the vertical direction after
the top surface has moved beyond one scale height. 
Several streamlines of the flow are represented by dashed lines. 
The elongated bubbles
exhibit a striking resemblance to the observed W4 superbubble
in the region above the star cluster. Figure 2
shows an \HI map of NTD and Figure 3 shows the H$\alpha$ map of DTS, both
overlaid with the best fit Kompaneets model, determined as follows. 
The best fit is determined
by finding the value of $\yt$ [in the full range (0,2)] for which the aspect 
ratio matches that of the observed superbubble. The H$\alpha$ map
of DTS reveals the ionized wall of the superbubble, and shows that it
reaches a maximum diameter of 3.6\arcdeg~ at latitude 4\arcdeg~
and thereafter starts to close. Figure 3 shows that there is a left-right 
asymmetry in the structure of the \halpha shell.
The maximum radius of the shell, measured relative to the longitude
of the cluster (134.7\arcdeg), is greater towards the right (2.1\arcdeg)
than towards the left (1.5\arcdeg); also, the shell is brighter
towards the left. This is consistent with the interpretation that
the bubble has blown into a denser medium on the left side, so that
it has not expanded as far. When ionization bounded (see \S\ 3.2), a shell
is brighter when closer to the cluster.
A slightly lower density toward the right-hand side may also explain
why the shell appears to extend to somewhat greater height on that side.
Since our intention is to model the mean atmosphere, and not such details
as a left-right asymmetry, we take an average value of 1.8\arcdeg~
for the maximum radius, thereby finding the Kompaneets model whose diameter
matches that of the superbubble. In the same spirit, the top of the
shell, while not clearly visible in the \halpha map, is estimated to be 
at latitude 6.9\arcdeg~, based on the shell structure on both left and right.
The star cluster is located at 0.9\arcdeg~ latitude.
Using this information, equations (\ref{z12}) and (\ref{rmax}),
which give expressions for $z_1$ ($z_2$), the distance from the star
cluster to the top (bottom) of the bubble, and $r_{\rm max}$, the maximum
half-width of the bubble,
are used to find the evolutionary stage of the bubble. Equating
the theoretical and observed aspect ratios yields
\beq
\frac{z_1}{r_{\rm max}} = -\frac{\ln(1-\yt/2)}{\arcsin(\yt/2)} 
\simeq \frac{6}{1.8} = 3.33.
\eeq
The equality is satisfied when $\yt = 1.9837$. In this calculation, 
we have assumed that the superbubble lies in a plane perpendicular to 
our line of sight.
Our value of $\yt$ is also consistent with the location of the bottom
of the bubble, since the Kompaneets model predicts that 
$|z_2|/z_1 = 0.14$ when $\yt=1.9837$, whereas the observations yield
$|z_2|/z_1 = 0.15$. The purely exponential atmosphere might be less 
realistic below the cluster, and finite external pressure
might play a role there if anywhere, so it is fortunate that 
the model approximately fits the 
bubble here as well (see \S 4 for further discussion of the physics
near the base of the bubble).

In addition to matching the aspect ratio, the Kompaneets model also 
yields a value for the mean scale height $H$ in the atmosphere if
we know the size of the superbubble. This requires a distance,
which we take to be 2.35 kpc, based on the distance to OCl 352 estimated
by Massey et al. (1995). At this distance,
the H$\alpha$ map yields $z_1 \simeq 246$ pc and $r_{\rm max} \simeq 74$ pc.
Since $\yt=1.9837$ in the Kompaneets model implies that $z_1 =9.62\, H$
and $r_{\rm max} =  2.89\, H$, matching the 
Kompaneets model to the observations requires 
\beq
H \simeq 25 \, {\rm pc}.
\eeq
Even though the exact position of $z_1$ is is not very clear 
in the H$\alpha$ map, the unmistakable presence of a maximum radius 
$r_{\rm max}$ alone can be used to obtain a rough estimate for $H$, 
since the highly elongated nature of the
bubble implies that $r_{\rm max}$ ($\simeq 74$ pc) must be close to 
the limiting theoretical value $\pi H$.
Furthermore, although our estimate for $H$ is based on the
Kompaneets model, it is worth pointing out that a scale height of this
approximate magnitude is unavoidable in any model, since the current 
radius of the bubble (maximum value $\simeq 74$ pc) must be significantly 
greater than $H$ in order for the bubble to have become so elongated.

\subsubsection{The Swept-up Mass in the Shell}

\label{sec_swept}

The mass distribution in the shell is an important factor in determining
the ionization structure of the atmosphere, which is calculated
in \S~3.
The spatial profiles can be used to find the streamlines of the flow, 
since the latter are normal to the local surface at all stages of evolution
(see Figure 1). Subsequently, one can integrate along the streamlines
in the exponential atmosphere to obtain the swept-up surface (column) density
at all points on the shell. Figure 4 shows the dimensionless surface 
density of the swept-up shell
$\tilde{N}$ (unit $n_0 H$) when $\yt=1.98$ (the stage of
evolution which fits the W4 superbubble) as a function of (a) height
$z/H$, and (b) elevation angle $\theta$
(in degrees) measured relative to the $r$-axis. The surface density
of the Kompaneets shell is shown as a solid line, while the dashed
line represents the value for expansion along a straight-line path 
to the radius $s = \sqrt{r^2 + z^2}$
through the same exponential atmosphere. At $\theta = 0\arcdeg$,
the radial (straight-line) expansion would yield a surface density $N_s
= n_0 r(z=0)/3 \simeq 2 n_0 H/3$, since $r(z=0) \simeq 2H$ at the advanced
stages of evolution of a Kompaneets bubble. The Kompaneets streamlines
yield a lower $N$ than radial expansion for all angles $\theta > 0\arcdeg$,
since the streamlines tend to be deflected downwards, and yield higher 
values for most angles $\theta < 0\arcdeg$ for the same reason.
The crossover between the two curves occurs just below $\theta = 0\arcdeg$.
At $\theta=0\arcdeg$, the Kompaneets streamlines yield $N = 0.67 \, n_0 H$, 
versus $N_s = 0.69 \, n_0 H$ for radial streamlines.
The Kompaneets streamlines yield values of $N$ up to about 15\% 
higher than the radial value near the bottom of the bubble, but is
more than an order of magnitude below the radial value near 
the top. The continual downward deflection of streamlines 
means that most of the mass is not transported very far upwards, if
at all. In the late stages of the Kompaneets model,
about 55\% of the matter in the shell is found below $z=0$, about
75\% is found below $z=H$, and nearly 90\% below $z=2H$.

\subsubsection{Age of the W4 Superbubble}

The time-dependent solution (\S\ \ref{apdx_time}) allows us to make an age estimate 
for any bubble if we know the stage of evolution (represented by $\yt$). For
application to the W4 superbubble, we note that the dimensionless time
$\ttil=6.3$ when $\yt=1.98$.
Converting to dimensional form, we obtain
\beq
\label{age}
t = 1.18 \left( \frac{n_0}{1 \,{\rm cm}^{-3}} \right)^{1/3} \;
\left( \frac{H}{25 \,{\rm pc}} \right)^{5/3} \;
\left( \frac{3 \times 10^{37} \,{\rm ergs}\,{\rm s}^{-1}}{L_0} \right)^{1/3} \;
{\rm Myr}.
\eeq
Although $L_0$ can be estimated from direct observations of the stellar
winds in the cluster stars or from the spectral type of the stars (NTD),
and $H$ is estimated from the size and shape of the bubble, 
a final estimate for the bubble age requires a constraint on $n_0$,
which we obtain in \S 3 from the ionization structure
of the atmosphere.

\section{Ionization Structure}

Along with the powerful winds which drive the W4 superbubble, the nine
O stars also produce an extremely strong ultraviolet radiation field.
The dynamics of the superbubble
is dominated by the energetic winds, but the optical and radio continuum
appearance of the region is determined by these ionizing photons.

The evolution of \HII regions without powerful winds has been well studied
(e.g., Spitzer 1978; Tenorio-Tagle 1982)
and found to be incapable of producing extreme structures such as the 
W4 superbubble.
The strongest constraint on dynamical \HII regions is the inability to move
large amounts of material at speeds much faster than the sound speed 
$c_{\rm s}$ in the hot, ionized gas of temperature $T \sim 10^4\,$K.  
Unlike the wind-blown superbubble,
where $T \sim 10^6\,$K, an evolving
\HII region cannot expand more than a few tens of parsecs over its million
year lifetime. The ionization structure of a wind-blown region is, however,
produced by the irradiation of the swept-out cavity by the ionizing UV
photons from the O stars. This allows observations of the ionized hydrogen 
to be used as a probe of the ambient atmospheric conditions.

\subsection{The Formation of \HII Regions Without Powerful Winds}

Before calculating the ionization structure of the wind-blown bubble,
it is relevant to review the ionization structure of \HII regions in 
a variety of media. 

\subsubsection{Uniform Medium}

Within a constant density environment, the ionizing
photon flux $\Phi_*$ from the central OB association produces a 
classical Str\"omgren sphere \HII region with radius 
\beq
\label{e_Rs}
\Rst = \left( {3 \Phi_* \over 4 \pi\,n_0^2\,\alpha_B} \right)^{1/3}.
\eeq
In the above, $n_0$ is the number density of hydrogen nuclei inside the
Str\"omgren sphere and $\alpha_B \simeq 2.6 \times 10^{-13}$ cm$^3$ s$^{-1}$
is the recombination coefficient for hydrogen.
For the nine O stars at the center of the W4 superbubble,
$\Phi_* \simeq 2.3\times 10^{50}\,{\rm s}^{-1}$ (DTS), 
and the Str\"omgren radius is
\beq
\label{e_Rsw4}
\Rst = 200 \left( {1 \,{\rm cm}^{-3} \over n_0} \right)^{2/3}\;
\left( {\Phi_* \over 10^{50}\, {\rm s}^{-1}} \right)^{1/3} \; {\rm pc}.
\eeq

Once formed, the initial \HII region expands due to the overpressure of the 
hot, ionized gas relative to the external neutral gas. 

\subsubsection{Exponentially Stratified Medium}

A second useful example to consider is the instantaneous structure of 
the initial \HII region in an exponentially stratified atmosphere 
with scale height $H$. 
Because the density drops precipitously with increasing $z$, it is 
possible that in the upward direction the \HII region will be unbounded. 
If $H$ is much larger than $\Rst$, so that the density gradient across the \HII
region is small, a quasi-spherical nebula is produced; however, if $H$ is
much smaller than $\Rst$, the ionizing photons are free to escape the
stratified ISM in the upward direction.  Neglecting the diffuse ionizing
field, the critical ratio 
$(\Rst/H)_{\rm crit}$ at which photons traveling directly upward
can escape is found by equating the total number of recombinations within an 
infinitesimal solid angle $d\Omega$ in the $z$ direction with 
the UV flux $\Phi_*$:
\beq
\label{e_H}
\Phi_*\, \frac{d\Omega}{4\,\pi} = 
\int^{\infty}_0 n_0^2 \exp( -2\,z/H) \alpha_B\,z^2\, dz\, d\Omega =
\frac{1}{4}\,n_0^2\,\alpha_B\,H^3\,d\Omega.
\eeq
Substituting equation (\ref{e_Rs}) into equation (\ref{e_H}) yields
$(\Rst/H)_{\rm crit} = \left( 3/4 \right)^{1/3} \simeq 0.9$.  
Thus, the condition for the initial Str\"omgren region to breakout is 
similar to that for the expansion of the wind-blown bubble, i.e., the  
characteristic radius must exceed the scale height.  

Figure 5 plots the 
instantaneous ionization boundary for a variety of $\Rst/H$ values showing
both confined and breakout \HII regions. In each case, the boundary of the 
ionized region is calculated by equating photoionization and recombination
along each solid angle. Note the gradual elongation of the nebula in the
upward direction with increasing $\Rst/H$ and the shallow slope to the
lower cavity when breakout is achieved. 
Like the initial Str\"omgren sphere in a uniform medium, these ionized
regions represent the initial conditions for further dynamical evolution
of the (overpressured) ionized gas. 
In extremely high-density 
(low $\Rst$) or low scale height regions, the breakout solutions 
represent the initial conditions for champagne-flow \HII regions 
(Tenorio-Tagle 1982), in which the ensuing dynamical evolution of the 
ionized gas at $\sim 10$ km s$^{-1}$ is important.

When $\Rst/H > 0.9$, a substantial fraction of the
atmosphere above the OB association may be completely ionized and a fraction 
of the ionizing photons may escape to the Galactic halo. The elevation 
angle $\theta_{\rm c}$ above which photons escape to infinity is 
determined by evaluating an analogous relation to equation (\ref{e_H})
along each solid angle (see also Franco, Tenorio-Tagle, \& Bodenheimer 1990).
This yields
\beq
\label{critangle}
\sin \theta_{\rm c} = 0.9 \frac{H}{\Rst}\ \ \ 
{\rm for} \ \ \ \Rst/H > 0.9.
\eeq
When $\theta_{\rm c} < \pi/2$ ($\Rst/H > 0.9$), the total fraction 
of ionizing photons that escape is
\beq
\label{ionfrac}
f = \frac{1}{2}\, 
\left( 1 - \frac{3}{2}\,\sin \theta_{\rm c} + \frac{1}{2}\,
\sin^3 \theta_{\rm c} \right).
\eeq
Equation (\ref{ionfrac}) yields limiting values $f=0$ when 
$\theta_{\rm c} = \pi/2$ (no escape of ionizing photons), and $f = 1/2$ when 
$\theta_{\rm c} = 0$ (the escape of all upward traveling photons). 

In \S\ 2, the scale height near the W4 region was estimated to be 
$H \simeq 25\,$pc, so that the critical value is $\Rst = 22.5\,$pc. For the
W4 superbubble $\Phi_* = 2.3\times 10^{50}\,$s$^{-1}$, so that the
critical density is $n_0 = 40\,$cm$^{-3}$. Therefore it appears likely that
during the {\it initial} illumination of the stratified ambient ISM,
before any expansion via winds, extensive ionization would have occurred,
with ionizing photons escaping to heights far above the cluster. 
In fact, we estimate
$n_0 \simeq 10\,$cm$^{-3}$ below, which gives $\Rst/H \simeq 2.3$, 
$\theta_{\rm c} \simeq 23\arcdeg$, and $f \simeq 0.22$. We emphasize that this
is just the initial stage. Later, when the material is swept into a shell
by the wind, the escape fraction $f$ can change (see \S\ \ref{struc_ifront}).

\subsection{The Illumination of Wind-Swept Cavities}

Given the remarkable similarity in appearance between the W4 superbubble and
the wind-blown bubble model presented in the last section, we
consider the illumination of such a cavity by the ionizing photon flux from 
a central OB association.  

\subsubsection{Structure of the Ionization Front}

\label{struc_ifront}

As discussed in \S\ \ref{sec_earlyevol}, the superbubble is filled 
with a tenuous hot medium; 
this does not provide significant attenuation of the ultraviolet radiation
field.  The swept up gas from the bubble is deposited 
along the cavity walls and is 
ionized by the Lyman continuum photon flux from the central OB association.
The pressure within the superbubble is determined from the time-dependent
Kompaneets solution (\S\ \ref{apdx_time}). Since the pressure within 
the cavity is much larger than the pressure in the ambient stratified 
medium (even when photoheated to $10^4\,$K),
the swept up material, which is in approximate pressure equilibrium
with the interior of the superbubble, is squeezed to higher density
than the ambient material. The density within the 
ionized zone of the swept-up shell is calculated below. 
Owing to the $n^2$ dependence of the recombination rate,
an important property of a wind-blown shell is that it absorbs a greater 
number of ionizing photons than the equivalent amount of matter in the 
undisturbed medium. 
The fraction of the shell which is ionized is easily determined by balancing 
ionizing photons and recombinations along a path from the central OB 
association through the shell. 

The ionization front is located where the Lyman continuum photon flux
becomes negligible. We compute this as a function of elevation angle 
$\theta$. For each $\theta$ three possible ionization states exist:
1) The ionization front may occur within the high pressured swept-up shell, 
in which case a very dense but cold neutral hydrogen (possibly molecular)
outer shell surrounds the ionized material. 2) The Lyman continuum flux may
escape the swept-up shell, only to produce an ionization front further out 
in the undisturbed stratified ISM. 3) The entire column along $\theta$ 
may be ionized, releasing Lyman continuum photons far beyond the bubble. 
State 2 is fairly rare in our model, because the swept-up material is 
very dense compared with the ambient exponentially stratified ISM,
so that if Lyman continuum photons penetrate through the shell, they will
likely break out of the atmosphere.

To quantify the above discussion, consider a segment of the shell located at 
an angle $\theta$ and distance $s$ with respect to the central OB 
association. The shape of the bubble is completely specified in terms of 
the dimensionless parameter $\yt$. 
The surface density $N (\theta, y)$ within this shell segment can be computed 
as described in \S\ \ref{sec_swept} and is proportional to $n_0\,H$.
We take the pressure within the shell to be $P_s(y) = 2\, P(y)$, where 
$P(y)$ is the average pressure in the bubble computed from the Kompaneets 
solution.  This is a reasonable estimate when using a thin-shell approximation
(see Bisnovatyi-Kogan \& Silich 1995, p.\ 670). The number density of hydrogen
nuclei in the shell is therefore
\beq
\label{nII}
n_{II} = {P_s(y)\over 2\,k\,T_{II}}.
\eeq
Dimensional values for $n_{II}$ can be obtained using the dimensional
pressure in the shell $P_s = 2 \tilde{P}\, (\rho_0 L_0^2/H^4)^{1/3}$, where
$\tilde{P}$ has a lower bound $\simeq 0.02$
at late times (see discussion in \S\ \ref{apdx_time}). For example, 
using $T_{II} = 8000$ K and $n_0 = 10$ cm$^{-3}$ 
(estimated below), we find that $n_{II} = 14.5$ cm$^{-3}$ at late times.
The maximum thickness of the ionized shell perpendicular to the local
tangent is
\beq
\label{maxthick}
\Delta R = N (\theta, y)/n_{II}.
\eeq
For an ionizing flux $\Phi_*$, the shell is ionized along the line-of-sight
from cluster center to a depth approximated by
\beq
\label{dels}
\Delta s \simeq 
\left( {\Phi_* \over 4\, \pi\, n_{II}^2\,\alpha_B\,s^2 }\right).
\eeq
The ionization depth perpendicular to the shell tangent, $\Delta s_R$,
requires multiplication of equation (\ref{dels}) by $\cos \phi$ where $\phi$ 
is the angle between the line-of-sight direction and the surface normal, 
yielding
\beq
\label{dsbydr}
{\Delta s_R \over \Delta R} \simeq 
\left( {\Phi_*\, \cos \phi\, k\, T_{II} \over 
2\, \pi\, N_s\,P_s\,\alpha_B\,s^2 }\right).
\eeq
If $\Delta s_R > \Delta R$, the entire shell is ionized and the Lyman
continuum photon flux illuminates the ambient stratified ISM outside the
bubble. If $\Delta s_R < \Delta R$, the ionization front occurs within 
the shell and a
much denser, cold shell layer in approximate pressure equilibrium overlies 
the ionized layer. Note that raising the ionizing flux, lowering the surface 
density, and lowering the pressure inside the bubble enhance the ability of 
Lyman continuum photons to penetrate through the shell. Determination of the 
location of the ionization front, if it exists outside the shell, is a 
straightforward adaptation of the method described in \S\ 3.1.

The location of the ionization front determines the appearance of the 
superbubble in the \HI observations.  If the ionization front is always
confined within the swept-up shell, then the region should appear as
a closed bubble in the \HI channel maps.  However, if the Lyman 
continuum photons
escape the shell along lines-of-sight above a critical $\theta$, then the 
ambient medium within a cone of angle $\pi/2 - \theta$ around $\hat z$ will be 
entirely ionized, and the region will appear as a ``chimney" in the 
\HI channel maps.  \halpha observations complement
the \HI channel maps by highlighting the dense ionized gas within the shell
even if the ionizing photons have penetrated into the low density medium
beyond it.
Thus, the different morphologies for the W4 superbubble described  in NTD 
(\ion{H}{1}: chimney) and DTS (\mbox{H$\alpha$}: closed bubble) are 
reconciled if the Lyman continuum flux breaks out toward the pole.

The height at which breakout of the Lyman continuum photons from the
dense shell can occur depends on the amount of material which has been swept 
up and on the pressure in the bubble, which sets the density within the 
ionized component of the shell.  In \S\ \ref{sec_swept},
the surface density of the shell was shown to vary dramatically from the
base of the bubble to the top, decreasing faster than the increase in
surface area (which varies as $s^2$) because of curved streamlines. 
Thus, breakout is expected to occur first at the top.
Although such quantities as the pressure and age of the bubble, and the 
surface density of the shell can be calculated in dimensionless form, where the
units are the appropriate combination of $n_0$, $L_0$, and $H$, no 
dimensionless model exists when a fourth parameter $\Phi_*$ is included. 
Therefore, specific dimensional models must be computed and analyzed. 
However, all the
relevant variables except the number density are reasonably determined for the
W4 superbubble, requiring only a one-dimensional search.  
The opening angle 
of the chimney observed in the \HI channel maps is the basis for 
determination of $n_0$. Figure 6 shows the location of the ionization 
front compared with the inner edge of the cavity for various values
of $n_0$. First, we estimate analytically the value of $n_0$ at which 
the ionization front becomes trapped in the shell near $z=0$.
The trapping condition, $\Delta s_R < \Delta R$, can be rewritten using
equation (\ref{dsbydr}) as
\beq
\label{trapping}
N \: > \: \frac{\Phi_* k T_{II}} {4 \pi s^2 \alpha_B P},
\eeq
where $P = 1/2\, P_s$ and $\cos \phi \simeq 1$. Furthermore, on dimensional 
grounds one expects that
\begin{eqnarray}
N & = & \tN\, n_0 H, \label{tildeN}\\
s & = & \ts\, H, \\
P & = & \tP \, (\rho_0 L_0^2 / H^4)^{1/3}. \label{tildeP}
\end{eqnarray}
The Kompaneets solution gives values for the dimensionless 
constants $\tN$, $\ts$,
and $\tP$ at the current epoch, namely $\tN \simeq 0.7$ at $z=0$, $\ts
\simeq 2$ at $z=0$, and $\tP \simeq 0.02$. Using these relations, equation
(\ref{trapping}) is transformed into the condition
\begin{eqnarray}
n_0 & > & \left( \frac{1}{\tN \tP \ts^2} \right)^{3/4} \;
\left( \frac{\Phi_* k T}{4 \pi \alpha_B} \right)^{3/4} \;
\left( \frac{1}{\mu m_{\rm H} L_0^2 H^5} \right)^{1/4} \nonumber \\
& = & 4.7 \; \left( \frac{0.02}{\tP}\right)^{3/4} \;
\left( \frac{0.7}{\tN}\right)^{3/4} \;
\left( \frac{2}{\ts}\right)^{3/2} \;
\left( \frac{\Phi_*}{2.3 \times 10^{50}\, {\rm s}^{-1}} \right)^{3/4} \nonumber \\
& & \times \; \left( \frac{T}{8000\, {\rm K}} \right)^{3/4} \;
\left( \frac{3 \times 10^{37}\, {\rm ergs}\,{\rm s}^{-1}} {L_0} \right)^{1/2} \;
\left( \frac{25\, {\rm pc}}{H} \right)^{5/4} \:  {\rm cm}^{-3}.
\label{n0min}
\end{eqnarray}
Figure 6, which is obtained using the same normalizing values as in 
equation (\ref{n0min}), also shows that the ionization front is 
trapped within the
shell at $z=0$ when $n = 5$ cm$^{-3}$. As $n_0$ is 
increased further, the ionization front gets trapped to progressively
greater heights. For fixed $n_0$, the ionization front opens outward
in a cone-like fashion above a critical height, as neither the shell nor 
the low density medium
beyond it are able to confine the ionizing photons. Beyond a density
$n_0 \simeq 10$ cm$^{-3}$, this opening point of the ionization front
rises very slowly with increasing $n_0$, since most of the original mass is
deposited within a few scale heights of the cluster (see the streamlines 
in Figure 1). We find our best fit model to the W4 superbubble by picking
the model whose ionization front turnoff point coincides with the height
of significant drop off of \HI emission. 
For NTD, this occurs at about 2.9\arcdeg ($\gtrsim 100$  pc) 
above the cluster (see Figure 2). 
This opening point is 
bracketed by the $n_0 = 10$ cm$^{-3}$ and $n_0 = 15$ cm$^{-3}$ curves,
with $n_0 \simeq 10$ cm$^{-3}$ a reasonable best fit.
Further \HI observations
should be able to confirm whether the apparent drop off
in \HI emission continues at higher latitudes, which would be consistent 
with the idea that the ionizing photons have penetrated the shell at heights
$\gtrsim 100$ pc.
It is interesting to note that the density estimate $n_0 \simeq 5$ cm$^{-3}$ 
made by NTD by looking at \HI gas adjacent to the chimney and at the latitude
of the cluster is about the same as the minimum density allowed by our
model (eq. [\ref{n0min}]). Furthermore, it is in good agreement with
our derived most likely value $n_0 \simeq 10$ cm$^{-3}$ given the 
uncertainties in both the observations and in some of our model parameters. 

The fraction of Lyman continuum photons which escape the superbubble
can be determined from these models and the analytic formulae
described in \S\ 3.1. For low density environments, little
material is swept up by the expanding bubble, and though the bubble appears
brighter than the background, the fraction of ultraviolet photons trapped by
the shell is insignificant; the escaping fraction may be deduced from the
arguments for a static exponential atmosphere, i.e., equations
(\ref{critangle}) and (\ref{ionfrac}).
However, when the density of the medium is large enough, the
exponential atmosphere becomes insignificant for this calculation
and the fraction of photons that
escape is determined by the
absorption within the illuminated shell. For our best fit model
($n_0 \simeq 10$ cm$^{-3}$), the fraction of solid angle
through which ionizing photons escape (when $\yt = 1.98$) is
approximately 15\%. At these opening angles, the shell is at a relatively
high latitude and contains little mass, so that nearly all the ionizing flux
is transmitted. Therefore, we calculate that a flux of ionizing photons
$\simeq 0.15\, \Phi_* \simeq 3\times 10^{49}$ s$^{-1}$ from the cluster OCl 352
escapes to higher latitudes.

Below $z \simeq 100$ pc, and particularly in the downward direction 
($z < 0$), this
model predicts that the stellar cluster does not completely ionize the dense
wind-swept shell.  There should then be a cooler, even more dense shell 
outside the ionization front, again in approximate pressure equilibrium.
It is therefore interesting that the IRAS observations show peak emission 
just outside the ionized shell along the lower part of the cavity 
(Heyer \& Terebey 1998). For maximum conversion of stellar into reradiated
infrared luminosity by dust, the shell should have substantial
optical depth. This can be checked by noting that the column density at
$z \simeq z_2$ is $\sim n_0\,H \simeq 10^{21}\,$cm$^{-2}$ (see Figure 4)
for our standard values of $n_0$ ($\simeq 10\,$cm$^{-3}$) and $H$ 
($\simeq 25$ pc). For a normal dust to gas ratio, this implies that
$A_V \sim 0.5$ mag, yielding significant self-obscuration
of our view into the lower bubble.  Accompanying this optical depth, 
we have the 
possibility of molecular gas (which might be swept up, having survived
the earlier evolution, or have reformed). 
Heyer \& Terebey (1998) also show CO emission
confined along a thin layer adjacent to the bottom edge of the superbubble,
indicating the presence of dense swept up molecular gas.

\subsubsection{Emission Measure}

\label{emiss_meas}

Considering the evolutionary scenario for a wind-blown superbubble outlined
in \S\ \ref{struc_ifront}, it is straightforward to produce models of the 
emission measure profile, $\int n^2 dl$, where the integral is
taken through the bubble along our line-of-sight, for a variety of $n_0$ 
values (Figure 7a-c). In each panel six epochs are shown corresponding 
to $\yt=0.5 - 1.98$ and the emission measure 
is presented as a grey scale, with darker regions corresponding 
to higher emission measure.  A cavity in \HI channel maps should occur
wherever emission is observed; in these regions the gas is ionized.
Each figure also shows the instantaneous location of the ionization front
separating the ionized from the neutral gas.
Thus, when $n_0$ in the ambient ISM is low, as in Figure 7a,
the region always appears as a giant cavity in \HI.  The ionization front
occurs in the ambient medium outside the shell, so that in
\halpha or radio continuum the region shows a thin-shelled bubble with
a very bright halo beneath. As $n_0$ is increased, the \HI 
chimney narrows and the bright \halpha halo disappears from beneath the
bubble. In the highest density case presented, ($n_0 = 10$ cm$^{-3}$), 
the upper shell is opaque to the ionizing photons for a time,
once sufficient mass is swept up into it, but eventually becomes transparent
when the internal pressure (and the density of the shell) has dropped
sufficiently.

The \halpha shell thickness observed by DTS falls in the range 0.2\arcdeg~
to 0.5\arcdeg~ at all heights. Our model predicts a thick shell near the
bottom of the bubble, and an extremely thin shell (much thinner than observed
by DTS) in the upper portion of the bubble (Figures 7a-c). The former
is due to the large amount of mass swept up in the increasing exponential
atmosphere below $b=0.9\arcdeg$ ($z=0$), which is probably not a 
good representation of the
atmosphere there, and the latter is due to the uniform pressure assumed
in the bubble (see \S\ \ref{apdx_time} for further discussion). 
A much lower pressure
should exist near the top of the bubble at this stage, allowing for a 
thicker shell there. However, our model should give a reasonable estimate
for the shell thickness near $z=0$. The maximum shell thickness
there can be found by combining equations (\ref{nII}), (\ref{maxthick}), 
(\ref{tildeN}), and (\ref{tildeP}), yielding
\begin{eqnarray}
\Delta R & = & \frac{\tN}{\tP}\; \frac{n_0^{2/3} H^{7/3} k T}
{(\mu m_{\rm H})^{1/3} L_0^{2/3}} \nonumber \\
& = & 11.1  \left( \frac{\tN}{0.7} \right) 
\left( \frac{0.02}{\tP} \right) 
\left( \frac{n_0}{10\, {\rm cm}^{-3}} \right)^{2/3}
\left( \frac{H}{25\, {\rm pc}} \right)^{7/3} 
\left( \frac{3 \times 10^{37}\, {\rm ergs}\, {\rm s}^{-1}}{L_0} \right)^{2/3}
{\rm pc}. 
\end{eqnarray}
At a distance $d \simeq 2.35$ kpc, the standard value obtained above yields
\beq
\Delta \theta =  \frac{\Delta R}{d} \simeq 0.27\arcdeg,
\eeq
which falls within the range observed by DTS.

A direct comparison between the H$\alpha$ emission produced by the best-fit
model of the ionized wind-swept bubble ($n_0 = 10\,$cm$^{-3}$; 
last frame of Figure 7c) and 
the H$\alpha$ observation of DTS reveals the merits of the simulation. 
Conversion between emission measure and
rayleigh flux units (R) is straightforward: at $T_{II} = 8000\,$K, 
1 R = 2 pc cm$^{-6}$.  DTS compute an extinction at H$\alpha$ of
2.1 mag toward the star cluster which we apply as a uniform obscuring
medium to the model bubble. Also, as shown in Figure 8, the intensity profile 
of the DTS observation (averaged over 9 pixels or 9 pc) along the 
symmetry axis $r=0$ reveals a high background flux below the bubble and
a low background flux above the bubble (solid line). We empirically model this 
background as an exponential distribution with a characteristic scale of 
$3.3\arcdeg$ and a 140 R normalization at $b=0.9\arcdeg$ (the latitude of the 
stellar cluster) before attenuation (dashed line in Figure 8).
The best-fit model including the empirical background and attenuation to the
bubble is also shown in Figure 8 (thick line).  The overall profiles match 
remarkably well despite the many internal knots of emission appearing in 
the observations which have not been modeled. The brightness of the model
at the bottom of the shell ($z \lesssim -20$ pc) is significantly 
greater than the observed
\halpha flux. This deficit of observed \halpha flux was also noted by DTS,
and arises in comparison with any model which assumes that the bottom of the 
shell is ionization bounded, since the \halpha brightness 
is then determined by the ionizing flux.  A possible explanation for the
discrepancy is extinction due to dense swept-up neutral gas outside 
the ionization front ($A_V \sim 0.5$ mag; see earlier discussion).
However, the edge-brightening enhancement at the bottom
shell appears reasonable as does the slow decline
in brightness with height.  The intensity declines in two stages, separated
at $z \simeq 100\,$ pc, which marks the position within the model where the
ionization front breaks through the wind-swept shell. Both components
appear visible in the H$\alpha$ observations. 

In Figure 9, an intensity
profile is produced across the bubble at a height $z = 50\,$pc. The
model has been attenuated in the same manner as discussed for Figure 8
(at this height the background flux is 100 R in the empirical model).
The H$\alpha$ data reveals enhanced emission within the bubble comparable
to the level expected from the model, but the two wall features are not
observed.  An intensity peak of the correct magnitude is observed at
$r = -40\,$pc, and is consistent with the fact that the \halpha shell is
pinched at low latitudes (Figure 3). As noted above, the thickness of the
observed peak is greater than derived in the model, indicating that at this
latitude the pressure in the bubble may not be as high as expected 
from a simple Kompaneets solution.
The bubble edge is not apparent on the right hand side; however,
the intensity does decrease significantly where the edge should be.
Part of the difficulty in matching the edge-brightened intensities may
be the requirement, satisfied in the model but perhaps not in reality,
of significant coherence along our line-of-sight. If the thin bubble
wall has non-axisymmetric structure, then the intense edge emission will
be spread over a larger area and the bright cusp will not be observed.

\section{Discussion}

Given the best fit parameters $n_0 \simeq 10$ cm$^{-3}$, $L_0 \simeq 
3 \times 10^{37}$ ergs s$^{-1}$, and $H \simeq 25$ pc for the
W4 superbubble, equation (\ref{age}) yields the best-fit
age estimate 
\beq
\label{finalage}
t \simeq 2.5 \; {\rm Myr}.
\eeq
This is less than the estimated main-sequence lifetime (3.7 - 4.3 Myr)
of the O stars in the cluster (NTD), and is therefore consistent with NTD's
interpretation that the superbubble is blown by the combined winds
of the young cluster in which no supernovae have yet occurred.
The cluster age has also been estimated by various authors to be 
$\leq 2.5$ Myr (see DTS).
The superbubble is on its way to blowing out of the Galactic disk unless
the atmospheric structure changes significantly at greater heights.
Given our best fit parameters, the blowout parameter (defined in \S\ 
\ref{sec_earlyevol})
$b \simeq 25$ (see eq. [\ref{blowout}]), so that use 
of the Kompaneets model is justified. 
A quick estimate of the superbubble age can also be made
using an important feature of expansion in a stratified atmosphere
pointed out in \S\ \ref{apdx_time};
the base radius at $z=0$, about 50 pc in this case, can be equated to
$\Rs$ in the spherical solution (eq. [\ref{Rs}]), and our best estimates 
for $n_0$ and $L_0$ used to obtain $t \simeq 2.1$ Myr. This is in reasonable
agreement with the value (eq. [\ref{finalage}]) determined directly from
the Kompaneets model.

The shape of the W4 superbubble provides an important surprise: the ambient
interstellar gas around the cluster OCl 352 was very strongly stratified
in the vertical direction. The estimated scale height $H \simeq 25$ pc
is based on the best fit Kompaneets profile and the estimated distance
2.35 kpc to the cluster. This is to be compared with typical estimates
$H \gtrsim 100$ pc for interstellar gas near the Galactic plane based on 
emission from \HI or CO. For example, the FCRAO CO survey of the outer Galaxy
(Heyer et al. 1998), where the Perseus arm lies,
finds an overall mean CO scale height of 113 pc\footnote{One should note that
this is based on kinematic distances determined by the standard
circular rotation curve of the Galaxy, which can overestimate the distance
for gas in a spiral arm, thereby also overestimating the scale height.
Heyer \& Terebey (1998) point out this effect; the kinematic distance to 
OCl 352, based on associated molecular gas, ranges from 3.5 to 4.0 kpc, while
the spectroscopic distance to OCl 352 is only 2.35 kpc (Massey et al. 1995).}.
Our scale height determination
is, in contrast, based on a dynamical model of an observed superbubble,
and is applicable only to the immediate vicinity of the stellar cluster
OCl 352 and the W4 \HII region. Our scale height estimate would increase 
linearly in proportion to the cluster distance if it were actually farther
than the 2.35 kpc recently determined by the spectroscopic parallax
method (Massey et al. 1995). 
However, we do not expect the actual distance to be much different, since
many previous estimates for the distance to OCl 352 or the W3/W4/W5 \HII
regions have, using a variety of techniques, yielded values in the range
1.6 to 2.7 kpc (see NTD). Similarly, the estimate $H \simeq 25$ pc cannot
be explained away due to any of the simplifications of the Kompaneets model.
All models for bubble expansion, whether analytic, semianalytic, or
numerical (e.g., Kompaneets 1960; MacLow \& McCray 1988; Tomisaka \& 
Ikeuchi 1986; MacLow et al. 1989),
show that the bubble remains nearly spherical as long as the ratio of
the radius in the plane $z=0$ to the scale height $R/H < 1$. Significant
elongation takes place only when $R/H \gtrsim 2$, i.e., the bubble has had a
chance to fully sample the stratified atmosphere. The W4 superbubble is 
currently highly elongated and has a radius in the plane $R \simeq 50$ pc, 
hence one can conclude that $H < 50$ pc, and most likely $H \approx 25$ pc,
using very simple physical arguments. 

The relatively small value of $H$ in comparison to the mean ISM value ($\gtrsim
100$ pc) implies that significant vertical compression of the interstellar
gas has taken place here, perhaps not such a surprising finding
for a star-forming region. Consistent with this, the most likely value 
for $n_0$ ($\simeq 10$ cm$^{-3}$) is also 
significantly greater than the mean ISM value ($\sim 1$ cm$^{-3}$).
It is interesting to note that the total surface density of gas,
which is proportional to $n_0 H$, is about a factor of two greater than
implied by the mean ISM values. This is consistent with the idea that some
lateral inflow of gas has also taken place during the assembling of the
molecular cloud.

The Kompaneets model idealizes the atmosphere as being purely exponential,
so it is all the more striking that the $\yt=1.98$ profile provides such a
good approximation to the overall shape of the superbubble (Figure 3). 
Thus, the 
exponential atmosphere and scale height $H \simeq 25$ pc should be reasonable
approximations for the mean atmosphere into which the superbubble has expanded.
At large heights, towards the top of the superbubble, it is possible that
the stratification is not so great. Although there is relatively little
mass there, it may be more than implied by an exponential atmosphere
with $H \simeq 25$ pc. 
Possible deviations from the assumed atmosphere, e.g., the
extent to which the medium with $H \simeq 25$ pc can be embedded in 
one with larger $H$ at high latitudes and/or adjacent longitudes, can be
best addressed with future modeling using the thin shell approximation or
numerical models. However, we note that various models of
bubble expansion (e.g., MacLow \& McCray 1988) have shown that a
bubble changes shape dramatically when it travels from a relatively low
scale height atmosphere (e.g., the \HI cloud layer) to one with
greater scale height (e.g., the \HI intercloud layer); that is, the
bubble radius expands to become comparable to the local scale height.
This sort of expansion has not yet happened to the W4 superbubble. 
Hence, we conclude that 
the effect of extra mass near the top of the bubble cannot be very great.

Below the star cluster, the density probably decreases eventually,
certainly on passage past the midplane of the Galaxy,
unlike the increasing exponential function in the Kompaneets model.
However, the presence of a decreasing density gradient in the downward 
direction must be much less pronounced than
that above the cluster, or else the superbubble
would blow out in both directions. The finite pressure of the external medium
may also play a role in confining the downward expansion.

Deviations from a pure exponential atmosphere are already evident in
Figures 2 and 3. Figure 3 reveals that the superbubble is narrowest at a 
latitude almost equal to that of the star cluster. Such a pinch is 
usually caused by a local maximum in the density (e.g., MacLow \& McCray 1988).
There is clear evidence for high density regions which inhibit the 
expansion immediately adjacent to the cluster; CO and IRAS images reveal
a dense cloud near the left (east) pinch and the right hand (west) side 
is certainly affected by the presence of the W3 molecular cloud.

Further evidence for a more complicated atmosphere is provided by
the \HI and CO maps (NTD; Heyer et al. 1996),
which reveal embedded clouds in the cavity.
These clouds are most likely remnants of an initially
clumpy medium; a superbubble can move around dense clumps and continue to
expand (MacLow \& McCray 1988). The CO maps reveal two cometary 
shaped clouds (Heyer et al. 1996), and NTD find a V-shaped filament in \HI
(visible in Figure 2), with the higher latitude CO cloud at its base.
The V-shaped ``streamers'' are suggestive of material being stripped
from the molecular cloud by a free-streaming wind, but Heyer et al. (1996)
argue that the cometary shape of the CO clouds is likely due to the
sculpting effects of the ionizing UV radiation field of the O stars.
Our model supports the view that the CO clouds are embedded
in a hot rarefied gas in which the ionizing UV radiation from the central
source propagates freely, i.e., region (b) in the description of a wind-blown
bubble given in \S\ \ref{sec_earlyevol}. The region (a) consisting of 
free-streaming wind 
remains relatively small during the evolution due to the high pressure
in the hot shocked region (b). We estimate the current radius of this 
region to be $R_1 \simeq 10$ pc (see Appendix B). For comparison, the bottom
of the lower latitude embedded CO cloud observed by Heyer et al. (1996)
is about 16.5 pc above the cluster.
We note that photoevaporation of these clouds will act to cool the 
shocked gas in the bubble
interior (McKee, Van Buren, \& Lazareff 1984) and slow down the expansion
of the bubble. This will tend to increase the inferred age of the 
bubble given by equations (\ref{age}) and (\ref{finalage}). 

The thickness of the ionized shell in our model agrees well with 
observations of the superbubble near $z=0$, as shown in \S\ \ref{emiss_meas}.
However, the shell thickness in our model becomes progressively thinner
at high latitudes than revealed by the \halpha map of DTS.
This can be due to several reasons. As explained in \S\ \ref{apdx_time}, 
our estimate of the pressure applies only near the base of the
bubble. In reality, a 
pressure gradient must be present in the bubble at this stage, so that
the pressure is much lower near the top, allowing for a thicker 
(but less bright) shell there.
Additionally, a swept up magnetic field may contribute significant pressure
within the shell, and prevent it from being compressed to the level in 
our model. A tangential magnetic field within the shell might also explain
another puzzling feature of the observed superbubble. The Kompaneets
model, and indeed all expansion models, predict that a bubble is in the
rapidly accelerating blowout phase by the time it is so elongated
(recall the decelerating expansion only lasts while the bubble is within
one or two scale heights above the plane, and still has a nearly
spherical shape). Hydrodynamic simulations of expanding superbubbles
show that the shell fragments during this stage due to a Rayleigh-Taylor 
instability (MacLow et al. 1989). Yet, the \halpha map of DTS shows that
the W4 supershell is relatively smooth. A plausible
explanation for the suppression of the Rayleigh-Taylor instability 
is the stabilizing effect of a swept-up tangential magnetic field
in the shell (e.g., the simulations of Stone \& Norman 1992).
We investigate this issue further in a future paper; preliminary results
can be found in Komljenovic, Basu, \& Johnstone (1999).

While magnetic fields might explain some
features of the W4 superbubble, we caution that they can also raise
complications. For example, a dynamically significant magnetic
field aligned predominantly parallel to the Galactic plane would 
prevent the bubble from attaining its present elongated shape, both by
potentially increasing the scale height $H$ and also by exerting a force
that inhibits expansion perpendicular to the plane. Indeed, simulations
of bubble expansion in a stratified medium with a dynamically
significant field parallel to the Galactic plane show that the bubble
expands preferentially {\it along} the Galactic plane (Tomisaka 1992), 
unlike the W4 superbubble.

Our model for the ionization structure of the bubble also predicts that a 
significant fraction ($\simeq 15$\%) of the ionizing UV photons from the 
cluster escape 
above the bubble. Although the top of the W4 superbubble is at a height
$z_1 \simeq 246$ pc, and we have no information about the density structure
above the superbubble, it is quite likely that if the ionizing photons
can penetrate the dense upper shell of the bubble, they will then 
continue to ionize the higher regions of lower density. 
The escape of ionizing photons is ultimately 
traced to the relatively low scale height $H$ of the local medium in comparison 
to the initial Str\"omgren sphere radius of the combined O stars.
If relatively low values of $H$, on the order of the value found in this
paper, are common around major star-forming regions and young star clusters, 
then the consequent penetration of ionizing UV photons far above the Galactic 
plane may be a natural explanation for the extensive presence and large
scale height ($>$ 1 kpc) of free electrons in our Galaxy 
(e.g., Reynolds 1989, 1991).

\section{Summary}

We have used a model for the dynamical expansion and ionization
structure of a superbubble to gain insight into ISM conditions
near the young cluster OCl 352. The Kompaneets solution provides
a reasonable fit to the observed shape of the \HI cavity and
\halpha shell of the W4 superbubble, and when combined with a distance 
to the cluster, yields an estimate for the mean scale height $H$
of the local atmosphere. Remarkably, we find that $H \simeq 25$ pc in 
this star-forming region, significantly less than the mean ISM value
near the Galactic plane. This relatively low
value is an unavoidable conclusion given the significant elongation
of the superbubble. The Kompaneets solution also allows us to estimate
the current pressure near the base of the superbubble. This estimate,
along with properties of the ionized shell of the superbubble, enable
us to constrain the value of the original ambient density $n_0$
near the cluster.
The total wind luminosity of the massive stars $L_0$
can be estimated from observations and/or the spectral type of the 
stars (Normandeau et al. 1996). Therefore, we can use the best values for
$n_0$ ($\simeq 10$ cm$^{-3}$), $L_0$ ($\simeq 3 \times 10^{37}$ 
ergs s$^{-1}$), 
and $H$ ($\simeq 25$ pc) to find a dimensional value for the age of the
superbubble. Our estimated age of 2.5 Myr is consistent with the
age of the cluster and the idea that the superbubble is blown by the
combined winds of the cluster stars, among which no supernovae have yet 
occurred.

We have calculated the ionization structure of a stratified medium, both
before and during the expansion of a superbubble within it. The combined
photoionizing ability of the nine O-type stars is sufficiently high and
the ambient scale height $H$ of the medium sufficiently low that initially
the ionizing photons escape to heights far above the cluster.
The expansion of the superbubble creates a dense shell, which could even
trap all the ionizing photons for a time. However, during the late stages,
when the bubble pressure (and therefore the shell density) has dropped
sufficiently, the ionizing photons penetrate beyond the upper portions
of the shell (of lower mass surface density) and escape
to greater heights; the shell still traps the ionizing photons
near the base ($z \lesssim 0$). This means that while the 
upper shell will be an enhanced region of \halpha emission, the superbubble
will have no visible lid in \HI. Thus, while the apparent closure of the
\halpha shell (Dennison et al. 1997) places a damper on the Galactic 
chimney hypothesis (at least at the present epoch), the W4 superbubble 
may continue to appear 
as a chimney when observed in \HI at higher latitudes than in the CGPS
Pilot project (Normandeau et al. 1996).
Finally, if the presence of a relatively low scale height
of gas is common near an OB cluster, the consequent escape of ionizing photons
may provide a natural explanation for the extensive presence of 
ionized gas in our Galaxy.

\acknowledgements

We thank Brian Dennison for making his \halpha data available to us.
This work was supported by the Natural Sciences and Engineering 
Research Council (NSERC) of Canada. D. J. was supported by an NSERC 
Postdoctoral Fellowship. S. B. was partially supported by the CGPS/CSP 
grant from NSERC.

\appendix

\vspace{0.5in}
\centerline{\bf APPENDICES}

\section{KOMPANEETS MODEL}

\subsection{Spatial Solution}

\label{apdx_spatial}

The Kompaneets model assumes that the energy source is located at $z=0$
in an exponential atmosphere (here the ISM) with density
\beq
\label{e_rhoH}
\rho(z) = \rho_0 \, \exp(-z/H),
\eeq
where $H$ is the scale height. Note that $z=0$ is not necessarily the
Galactic midplane.
The model assumes that the internal pressure of the bubble dominates any
external pressure: the atmosphere is pressure-free.
This means that all bubbles will continue to expand in the $z$ direction;
furthermore, the top of the shell will reach infinite $z$ in a finite time,
i.e., undergo blowout.
Physically, we can use the Kompaneets model when
we are confident that the external pressure $P_{\rm e}$ will not confine
the bubble before the upper shell begins to accelerate, i.e., the parameter
$b$ (eq. [\ref{blowout}]) must be significantly greater than one. 
This condition is certainly
satisfied by the W4 superbubble, as shown in \S\ \ref{sec_earlyevol}.

The internal pressure of the bubble is taken to be constant, and is
\beq
\label{presskomp}
P = (\gamma - 1)\, \frac{E_{\rm th}}{\Omega},
\eeq
where $\gamma$ is the ratio of specific heats, $E_{\rm th}$ is the thermal
energy of the bubble, and the volume $\Omega$ of the remnant is defined
in cylindrical symmetric coordinates $(r,z)$ by the integral
\beq
\Omega = \pi \int_{z_2}^{z_1} r^2(z,t) \, dz,
\eeq
in which $z_1$ and $z_2\, (< 0)$ are the top and bottom of the remnant, respectively.
The extremely high temperature of the shocked gas, and hence short
sound crossing time, is the reason that the internal
pressure is assumed to have readjusted to an isobaric state at
all times. Since the internal pressure dominates the external pressure,
the expansion speed is that given by the Hugoniot conditions for a
strong shock,
\beq
\label{vn}
v_n = \sqrt{ \frac{\gamma+1}{2}\,\frac{P(t)}{\rho(z)} }.
\eeq
The expansion is assumed to be directed normal to the local surface of the
remnant. In the Kompaneets approximation, the mass swept up by the outer shock
is assumed to reside in a thin shell behind the front, although the
effect of its inertia is not included in the model.

Using equation (\ref{vn}) and the assumption that the shock front moves
normal to itself everywhere, Kompaneets (1960) derived an equation
for the evolution of the shock front (see also Bisnovatyi-Kogan \& Silich
1995):
\beq
\label{kompode}
\left( \frac{\partial r}{\partial y} \right)^2
- \frac{\rho(z)}{\rho_0} \, \left[ \left( \frac{\partial r}{\partial z}
\right)^2 + 1 \right] = 0.
\eeq
In equation (\ref{kompode}), $y$ is a transformed variable (with
units of length) defined by
\beq
\label{ydefn}
y = \int_{0}^{t} \sqrt{\frac{\gamma^2-1}{2} \frac{E_{\rm th}}{\rho_0 \Omega} }
\, dt.
\eeq
The thermal energy $E_{\rm th}$ of the bubble is calculated from the
differential equation
\beq
\label{dedt}
\frac{dE_{\rm th}}{dt} = L_0 - P \frac{d\Omega}{dt}.
\eeq
This equation differs from that used in the blast wave formulation of the
problem, in which
$E_{\rm th}$ is a constant fraction of the energy deposited in an
initial blast.
Equation (\ref{dedt}) assumes that the wind luminosity is
thermalized at the inner shock front $R_1$, and that the only energy loss
is due to work done against the thin shell. The effect of radiative and
evaporative cooling of the bubble interior can also be added here if desired.

Kompaneets (1960) showed that equation (\ref{kompode}) could be solved
analytically by separation of variables, yielding the relation
\beq
\label{kompsoln}
r(z,y) = 2H \, \arccos \left[ \frac{1}{2} e^{z/2H} \, \left( 1 -
\frac{y^2}{4H^2} + e^{-z/H} \right) \right].
\eeq
Equation (\ref{kompsoln}) describes a sequence of shapes for the
shock front, which change as the parameter $y$ varies from 0 to
$2H$, at which time the top of the remnant formally reaches infinity.
Note that one can discuss the structure of the shock front 
without actually solving for the explicit time-dependence $y$ versus $t$
(\S\ \ref{apdx_time}) from equations (\ref{ydefn}) and (\ref{dedt}).
The top and bottom of the remnant, where $r=0$, are located at
\beq
\label{z12}
z_{1,2} = -2H \, \ln (1 \mp \frac{y}{2H}).
\eeq
The maximum radius of the bubble, where $\partial r/ \partial z = 0$, is
\beq
\label{rmax}
r_{\rm max} = 2H \, \arcsin \left( \frac{y}{2H} \right).
\eeq
The above equations show that by the time of blowout ($y=2H$), the
bubble has radius in the plane $r(z=0) = 2.09 H$, maximum radius
$r_{\rm max} = \pi H$, and has penetrated downward into the
atmosphere of exponentially increasing density to a location
$z_2 = -1.39 H$.

Equation (\ref{kompsoln}) illustrates an important property of the
Kompaneets model; the spatial solution is independent of the time
evolution. The shock front evolves through the sequence of shapes
given by equation (\ref{kompsoln}) which is a consequence only of the
atmospheric structure. However, the {\it rate} at which it evolves does
depend on the details of the energy input into the bubble, i.e., whether
the energy is input at one instant or in a continuous fashion, and
at what rate. Hence, the shape of an observed bubble
can supply information about the ambient atmosphere
independent of any knowledge of the energetics of the driving source.

\subsection{Time Evolution}

\label{apdx_time}

A numerical integration of equations (\ref{ydefn}) and (\ref{dedt}) yield
$y(t)$ and $E_{\rm th}(t)$, which implicitly give the time evolution
of the bubble. The solution for a wind-blown bubble in an exponential
atmosphere depends on only three parameters: the scale height $H$,
the density near the source $\rho_0$, and the wind luminosity $L_0$.
Hence, the problem is solved most naturally in a system of units
which are formed from these three parameters.
The unit of length is then $H$, the unit of mass is $\rho_0 H^3$,
and the unit of time is $(\rho_0 H^5/L_0)^{1/3}$. The units for various
physical quantities under this system are listed in Table 1.
We obtain the dimensionless solution by integrating the dimensionless
form of equations (\ref{ydefn}) and (\ref{dedt}). The dimensional
solution for any combination of $\rho_0$, $L_0$, and $H$ can be obtained
from this single solution using the units shown in Table 1.
In the following discussion, dimensionless variables are overlaid with a tilde
sign. Figure 10a shows the parameter $\yt$ versus time $\ttil$ in a solid line.
The integration is carried up to $\yt=1.98$, the stage at which the spatial
profile matches the W4 superbubble. We do not integrate all the way to
the limit $\yt=2$ due to the artificial nature of the solution at that time;
the top of the remnant reaches infinity at infinite speed and the
pressure drops to zero.
For comparison, the dimensionless radius of the spherical expansion
solution $\tilde{R}_{\rm s} = 0.76\, \ttil^{3/5}$ (from eq. [\ref{Rs}])
is plotted alongside in a dashed line. The relationship between
$y$ and $\Rs$ helps to assign some physical meaning to $y$.
For early times, when the expansion is nearly spherical,
inserting the spherical homogeneous solutions
for $\Rs$ (eq. [\ref{Rs}]) and $E_{\rm th}$ (eq. [\ref{Eth}]) into
equation (\ref{ydefn}) yields
\beq
\yt = 0.77\, \ttil^{3/5}
\eeq
if we take $\gamma = 5/3$. Therefore, $y$ is essentially equal to
$\Rs$ at early times. However, Figure 10a also shows that $y \simeq \Rs$
throughout most of the evolution. Although $y$ begins to fall below
$\Rs$ when the bubble has expanded beyond one scale height,
it is only 14\% below $\Rs$ when $\yt=1.98$. Figure 10b shows the
distance traveled by the bubble shell in three directions: 1) upward along
the $z$-axis, i.e., $\tilde{z}_1$, 2) downward along the $z$-axis,
i.e., $-\tilde{z}_2$,
and along the $r$-axis, i.e., $\tilde{r}(z=0)$. The dimensionless spherical
solution $\tilde{R}_{\rm s}$ is again given for comparison in a dashed line.
The top of the bubble $z_1$ eventually
moves much farther than $\Rs$ since it moves into
a progressively lower density environment, while the bottom
moves less far than $\Rs$ as it moves into a progressively
higher density environment.
With constant density along $z=0$, the time evolution of $r(z=0)$
closely follows that of $\Rs$,
even though the curved streamlines mean that $r(z=0)$ does not
follow a mass element in a Lagrangian fashion. There is an even
closer correspondence between $y$ and $r(z=0)$; even at the end of the
integration, the two values differ by only 4\%. Altogether, we find
the important relation
\beq
y \simeq r(z=0,y) \simeq \Rs.
\eeq
Therefore, the measurement of the radius $r(z=0)$ of any
observed bubble can yield quick estimates for $y$ and even the
age $t$ (if one has estimates for $L_0$ and $\rho_0$) since
$\Rs$ can be converted to a time through equation (\ref{Rs}).

The evolution of the thermal energy $\tilde{E}_{\rm th}(\ttil)$ is shown in
Figure 10c. For comparison, the dimensionless spherical solution
(from eq. [\ref{Eth}]) is
shown in dashed line. Interestingly, $E_{\rm th}(t)$ remains very close
to the linear time-dependence of the spherical solution even after
there is significant elongation
of the bubble. This means that the term $P d\Omega/dt$ in equation
(\ref{dedt}) remains nearly constant even when the bubble is becoming
elongated. At late times, $E_{\rm th}(t)$ does fall significantly
below the spherical expansion value and even begins to decrease.
This decrease is due to the rapid expansion of the bubble in the late
phases; $d\Omega/dt$ increases without bound, approaching an infinite
value at $\yt=2$, causing the second term in equation (\ref{dedt}) to
dominate the first, making $d E_{\rm th}/dt$ negative.
However, the Kompaneets approximation breaks down when the expansion
speed exceeds the internal sound speed, so we cannot trust that the
decrease of $E_{\rm th}(t)$ will occur in a more realistic model for
this stage.
Figure 10d shows the pressure $\tilde{P}$ versus $\ttil$. Again, the dashed line
shows the dimensionless spherical solution from equation (\ref{Press}).
As with other variables, the pressure follows the spherical solution
for much of the evolution, but it does drop significantly below it during
the late stages. This stage coincides with the rapid expansion of the
bubble, when one of the key assumptions of the Kompaneets approximation,
uniform pressure within the bubble, begins to break down.
Uniform pressure certainly cannot be maintained
when the expansion speed exceeds the internal sound speed. A pressure
gradient is then set up within the bubble. Hence, the pressure at the
base of the bubble, near the cluster, will not decrease to the values
shown in Figure 10d.

We estimate the current pressure
at the base by equating it to the mean pressure $P$ at a time when
rapid acceleration begins and the
expansion speed approaches the internal sound speed. Beyond
this critical time, we assume that the pressure at the base will not be able
to readjust to a new value in the short time available. This gives
us an estimate of the pressure at the base without going
through the complexities of a full numerical solution of the hydrodynamic
equations. We estimate the time at which the base pressure is
frozen at the mean value by comparing $dz_1/dt$ with the internal
sound speed $c_{\rm s,i} = \sqrt{kT_{\rm b}/\mu m_{\rm H}}$.
The bubble temperature $T_{\rm b}$ is estimated in Appendix B.
Figures 10e and 10f show the speed of the top boundary
$d\tilde{z}_1/d\ttil$ versus $\ttil$ and $\tilde{z}_1$ respectively;
the latter gives an idea
of where, rather than when, the acceleration phase occurs. Overplotted
on both curves, in dashed lines, is the internal sound speed
$\tilde{c}_{\rm s,i} = c_{\rm s,i}/[v]$, where $[v]$ is the
unit of velocity given in Table 1. The standard values $L_0 \simeq
3 \times 10^{37}$ ergs s$^{-1}$ and $H \simeq 25$ pc are used to
evaluate $[v]$, and
$n_0$ is taken to be 1 cm$^{-3}$. We use dimensional values
for $L_0$, $H$, and $n_0$ in this case since the temperature
depends implicitly on the thermal conductivity $C$ within the bubble.
Hence, the sound speed depends on a fourth independent parameter
and does not have a unique dimensionless value.
Figures 10e and 10f show that the
Kompaneets approximation breaks down after the top of the bubble departs
from the $t^{-2/5}$ deceleration of the spherical solution and goes into 
the acceleration phase. The crossing of the two curves takes place 
at dimensionless
time $\ttil = 4.9$ for $n_0 = 1$ cm$^{-3}$, when the dimensionless pressure is
$\tilde{P} \simeq 0.02$. Although this crossing would occur at a slightly later
time, and lower pressure, if we use a higher value for $n_0$, we use this
dimensionless critical pressure in our analysis, since the
combination of near-sonic velocity and very rapid acceleration make
pressure readjustment unlikely after this time even for somewhat higher
$n_0$ in the range 1 to 10 cm$^{-3}$. In \S\ 3, we use the lower limit
on the base pressure, $\tilde{P} \simeq 0.02$, as a parameter in our model
for the ionization structure of the atmosphere.

\section{SOME INTERIOR PROPERTIES OF THE W4 SUPERBUBBLE}

\subsection{The Interior Temperature}

The bubble temperature $T_{\rm b}$ is obtained from the model of
Weaver et al. (1977). Their equation (37) for the interior temperature $T$
in the bubble can be rewritten as
\beq
\label{Temp}
T = 63.6 \, \left( \frac{L_0}{\Rs} \right)^{2/7} (1 - r/\Rs)^{1/5} \: {\rm K}.
\eeq
It is valid for $r < \Rs$, where $\Rs$ is given by equation (\ref{Rs}).
Equation (\ref{Temp}) is obtained under the assumption of balance between the
outward thermal conduction flux and the inward mechanical energy flux due to
mass evaporation from the cool dense shell (see Weaver et al. 1977).
Though our bubble is highly elongated, equation (\ref{Temp}) should
give a reasonable estimate for the temperature near $z=0$ since
we find that the solution there is in many respects similar to the
spherical solution. Normalizing to the standard wind luminosity and
current radius $\Rs \simeq r(z=0) \simeq 50$ pc of the W4 superbubble,
we find a central temperature
(which is a representative value through most of a spherical bubble
due to the weak radial dependence of eq. [\ref{Temp}])
\beq
T_{\rm b} = 5 \times 10^6 \left( \frac{L_0}{3 \times 10^{37}\, {\rm ergs}\,
{\rm s}^{-1}} \right)^{2/7}
\left( \frac{50\,{\rm pc}}{\Rs} \right)^{2/7} \; {\rm K}.
\eeq

\subsection{The Inner Shock Radius}

We estimate the inner shock
front radius $R_1$ from the ram pressure of the free wind
\beq
P_{\rm w} = \rho_{\rm w} v_{\rm w}^2 = \frac{\dot{M}}{4 \pi r^2 v_{\rm w}}
v_{\rm w}^2 = \frac{(2 L_0 \dot{M})^{1/2}}{4 \pi r^2},
\label{Pw}
\eeq
where $\rho_{\rm w}$ is the wind density, $v_{\rm w}$ is the terminal
velocity of the wind, $\dot{M}$ is the
mass outflow rate, and the luminosity $L_0 = 1/2 \dot{M} v_{\rm w}^2$.
Equating this to the pressure $P_{\rm b}$ inside the region of hot shocked
gas yields a radius
\beq
R_1 = \frac{(2 L_0 \dot{M})^{1/4}}{(4 \pi P_{\rm b})^{1/2}}.
\label{R1}
\eeq
Now, the pressure in the bubble is written in dimensional form as
\beq
P_{\rm b} = \tP\, (\rho_0 L_0^2/H^4)^{1/3},
\eeq
where $\tP$ is determined from the time evolution of the Kompaneets
solution. In \S\ \ref{apdx_time} we estimated that $\tP \simeq 0.02$ 
at the current epoch, hence using this as well as the most likely values
$n_0 \simeq 10$ cm$^{-3}$, $L_0 \simeq 3 \times 10^{37}$ ergs s$^{-1}$, and
$H \simeq 25$ pc for the W4 superbubble, we find that
\beq
P_{\rm b}/k \simeq  1.1 \times 10^5 \; {\rm cm}^{-3}\, {\rm K},
\eeq
where $k$ is the Boltzmann constant. Using this value in equation (\ref{R1})
with a total mass loss rate for the cluster stars $\dot{M} \simeq 10^{-5} \;
M_{\sun}$ yr$^{-1}$ (see NTD), we find
\beq
R_1 \simeq 10 \; {\rm pc}.
\eeq
Hence, the hot shocked gas occupies nearly the entire volume of the
superbubble, and since $R_1 < H$, we are in retrospect justified in
using spherical geometry to calculate $R_1$.

\begin{figure}
\plotone{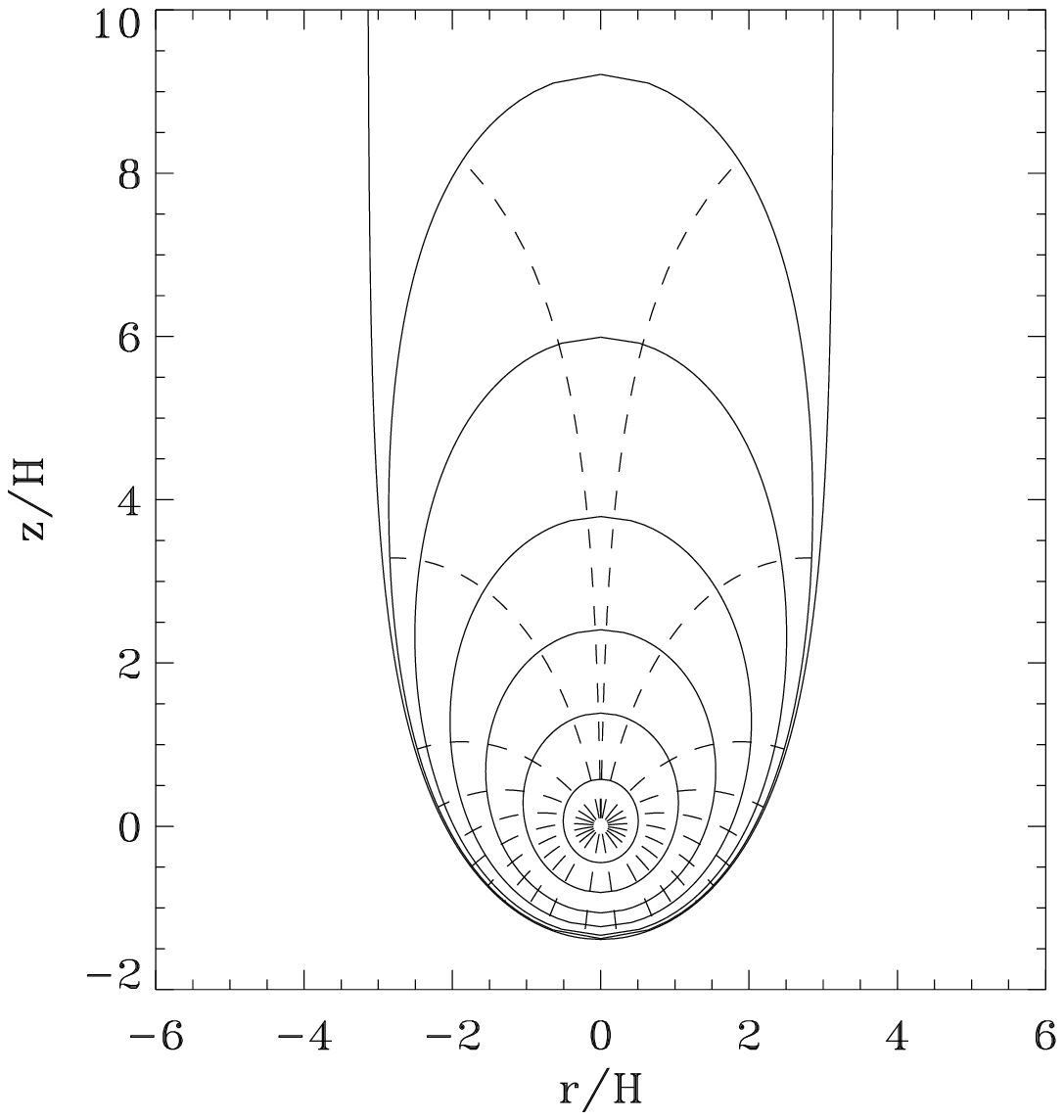}
\caption{Evolution of the shock front in the Kompaneets solution,
as the bubble expands in an exponential atmosphere $\rho = \rho_0 \exp (-z/H)$
and eventually blows out.
Solid lines show the position of the shock front at seven successive times
characterized by increasing values of the dimensionless parameter $\yt$
( = 0.5, 1.0, 1.4, 1.7, 1.9, 1.98, and 2.0). Dashed lines show the streamlines
of the flow for various points along the shock front.
}
\end{figure}

\clearpage 

\begin{figure}
\vspace{5in}
\includegraphics{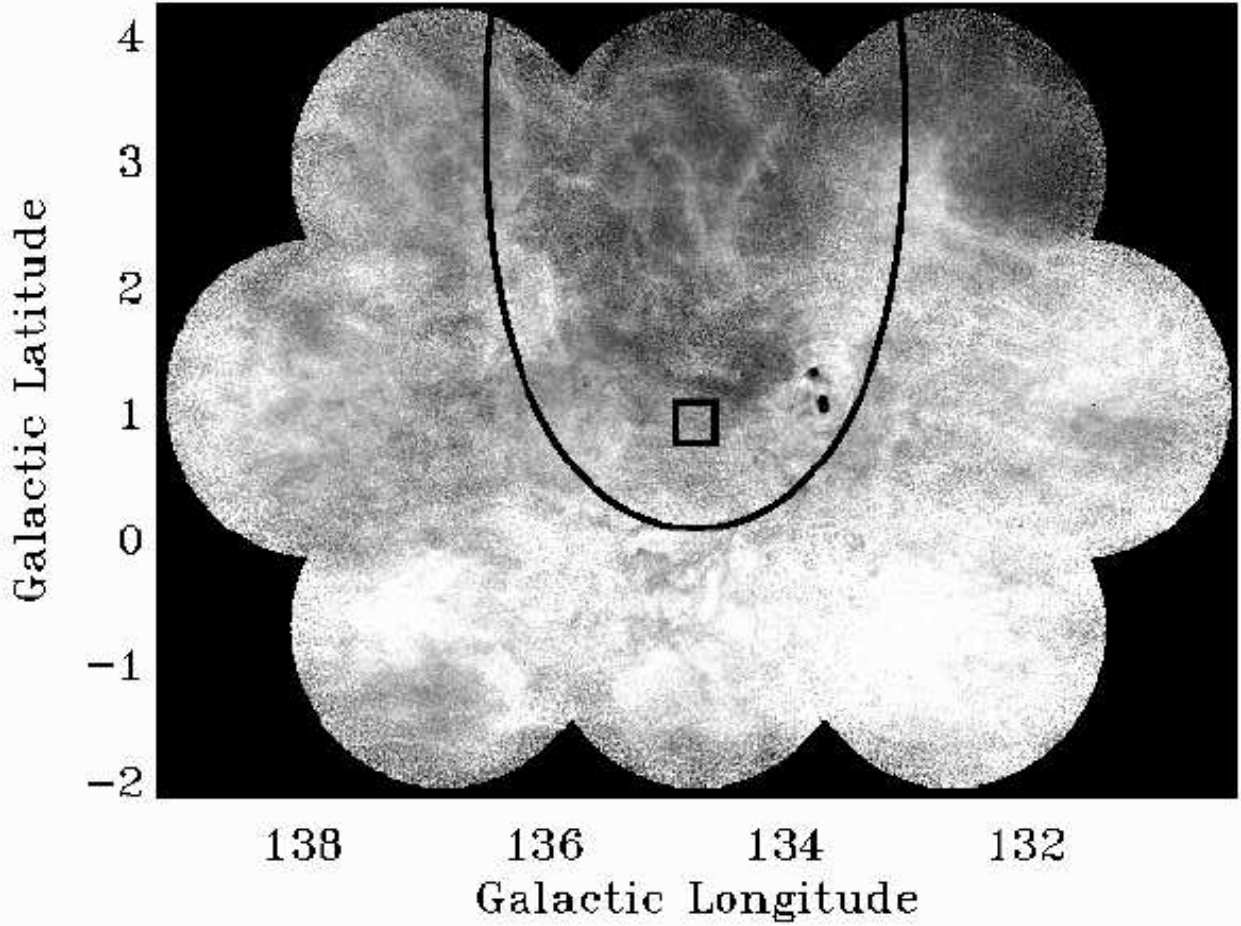}
\caption{\HI channel map showing the location of neutral
hydrogen in the vicinity of W4 and revealing the empty chimney discovered
by NTD. Lighter shadings correspond to brighter \HI emission. 
Overlaid on the figure is a Kompaneets solution with $\yt = 1.98$
and $H = 25\,$pc, centered on the star cluster OCl 352.
}
\end{figure}

\clearpage

\begin{figure}
\vspace{5in}
\includegraphics{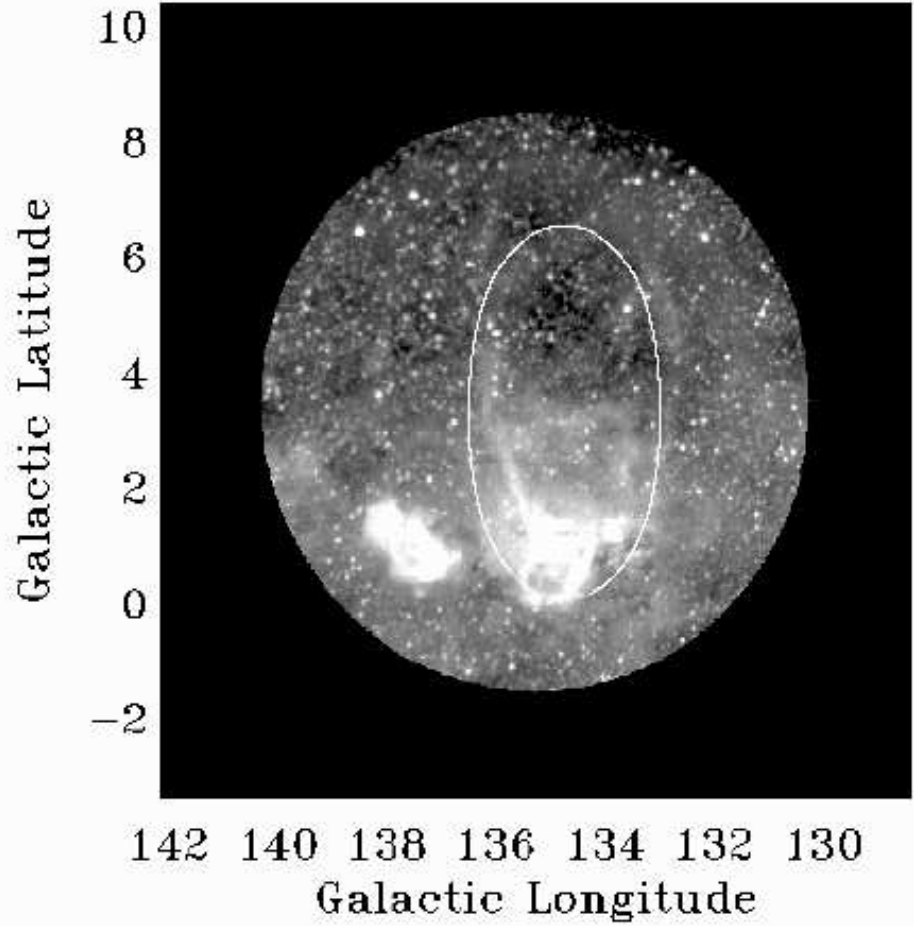}
\caption{\halpha map showing the location of ionized hydrogen
in the vicinity of W4 and revealing the extended superbubble discovered by
DTS. Lighter shadings correspond to brighter \halpha emission.
Overlaid is a Kompaneets solution with $\yt = 1.98$
and $H = 25\,$pc, centered on the star cluster OCl 352, as in Figure 2.
}
\end{figure}

\clearpage

\begin{figure}
\vspace{5in}
\includegraphics{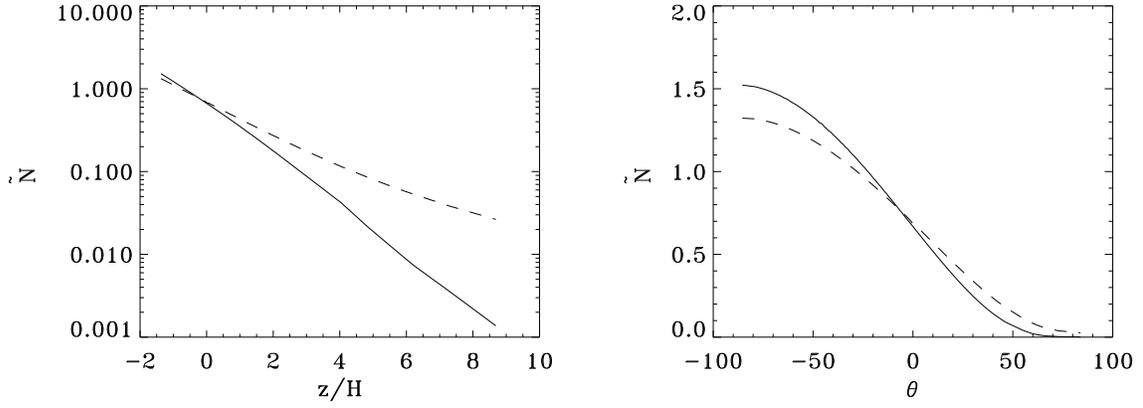}
\caption{Surface density $\tilde{N}$ (unit $n_0 H$)
in the swept up shell, when $\yt=1.98$, versus
(a) height $z/H$, and (b) angle $\theta$, in degrees, measured relative
to the $r$-axis ($z=0$). The dashed line shows the surface density if the
expansion had occurred along straight lines.
}
\end{figure}

\clearpage

\begin{figure}
\vspace{5in}
\includegraphics{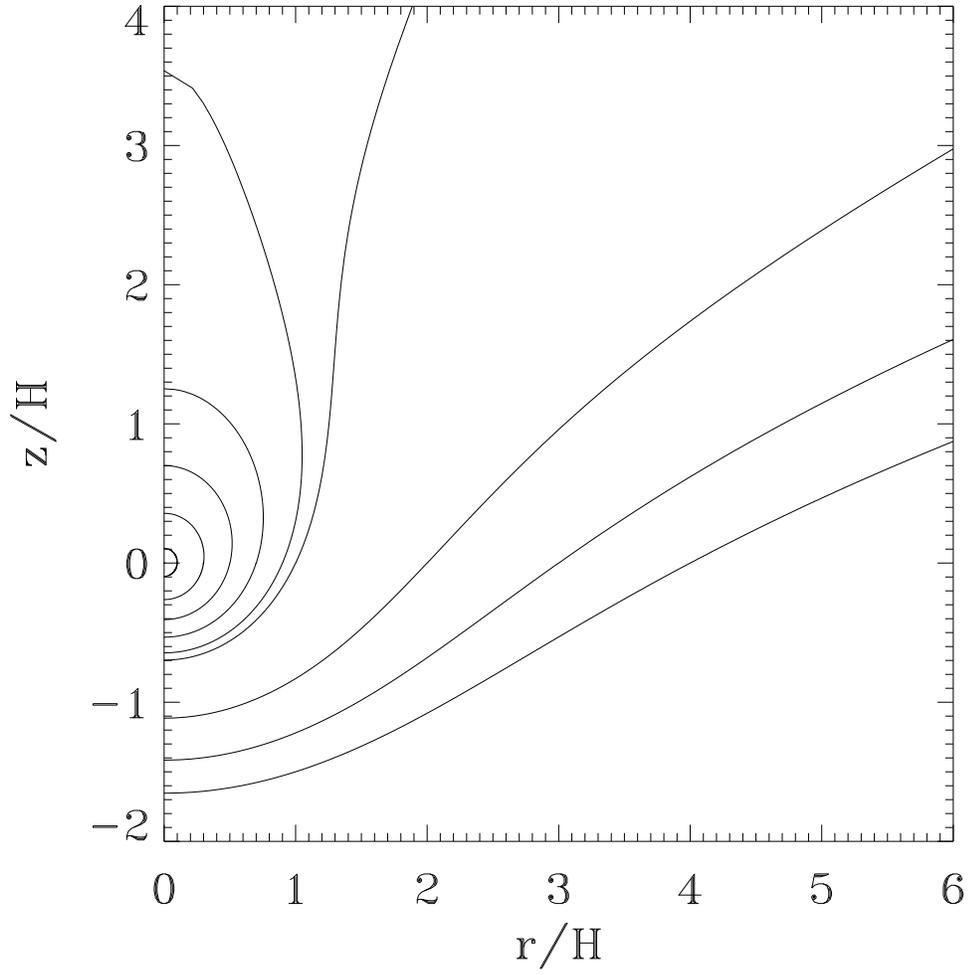}
\caption{Location of the initial ionization front around an \HII
region formed within an exponential atmosphere. The computed models, from
the innermost curve, are for $R_{\rm St}/H =$
0.1, 0.3, 0.5, 0.7, 0.9, 1.0, 2, 3, 4.
}
\end{figure}

\clearpage

\begin{figure}
\vspace{5in}
\includegraphics{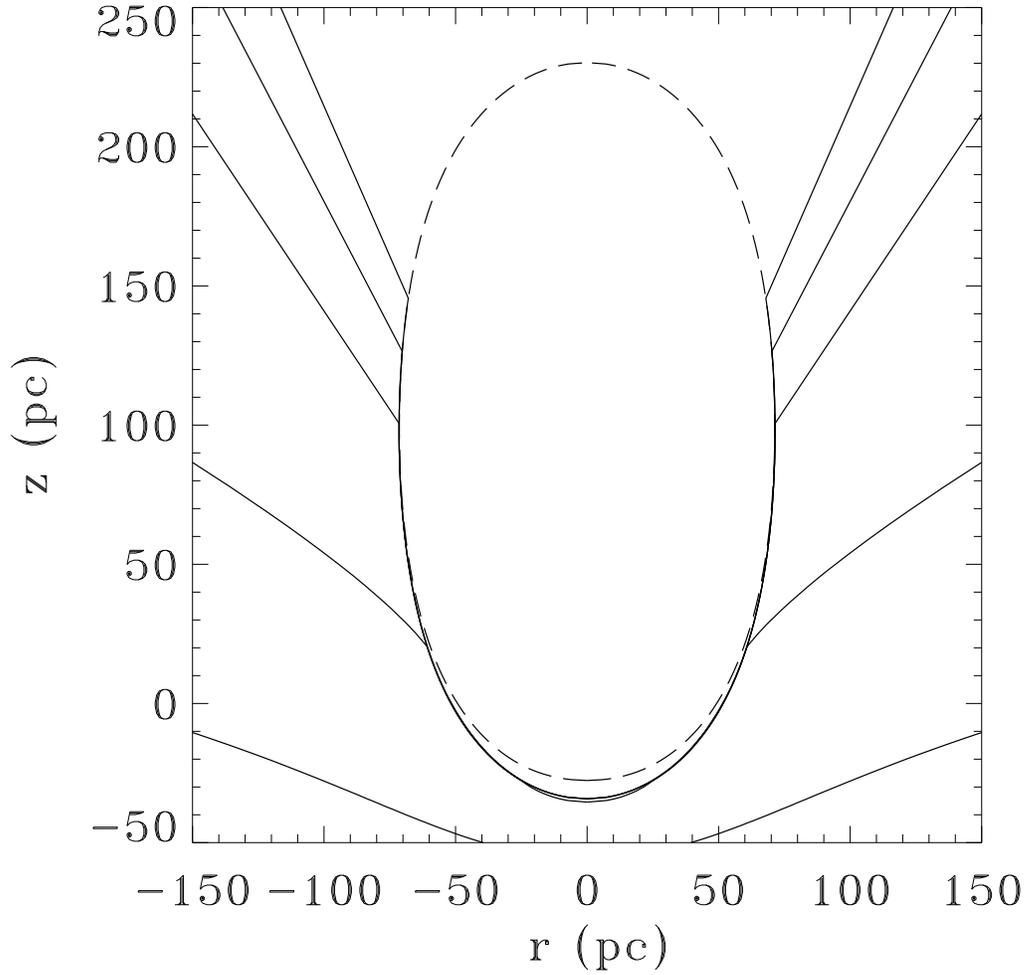}
\caption{Location of the ionization front around a wind-swept
bubble assuming the Kompaneets solution ($\yt=1.98$), the stellar
properties outlined in the text, and varying only  $n_0$,
the number density in the ionized exponential atmosphere ($H = 25\,$pc) at
$z = 0$. From bottom to top, $n_0 =$ 1, 5, 10, 15, 20 cm$^{-3}$.
The dashed line plots the inner boundary of the ionized wind-swept shell.
}
\end{figure}

\clearpage

\begin{figure}
\vspace{9in}
\includegraphics{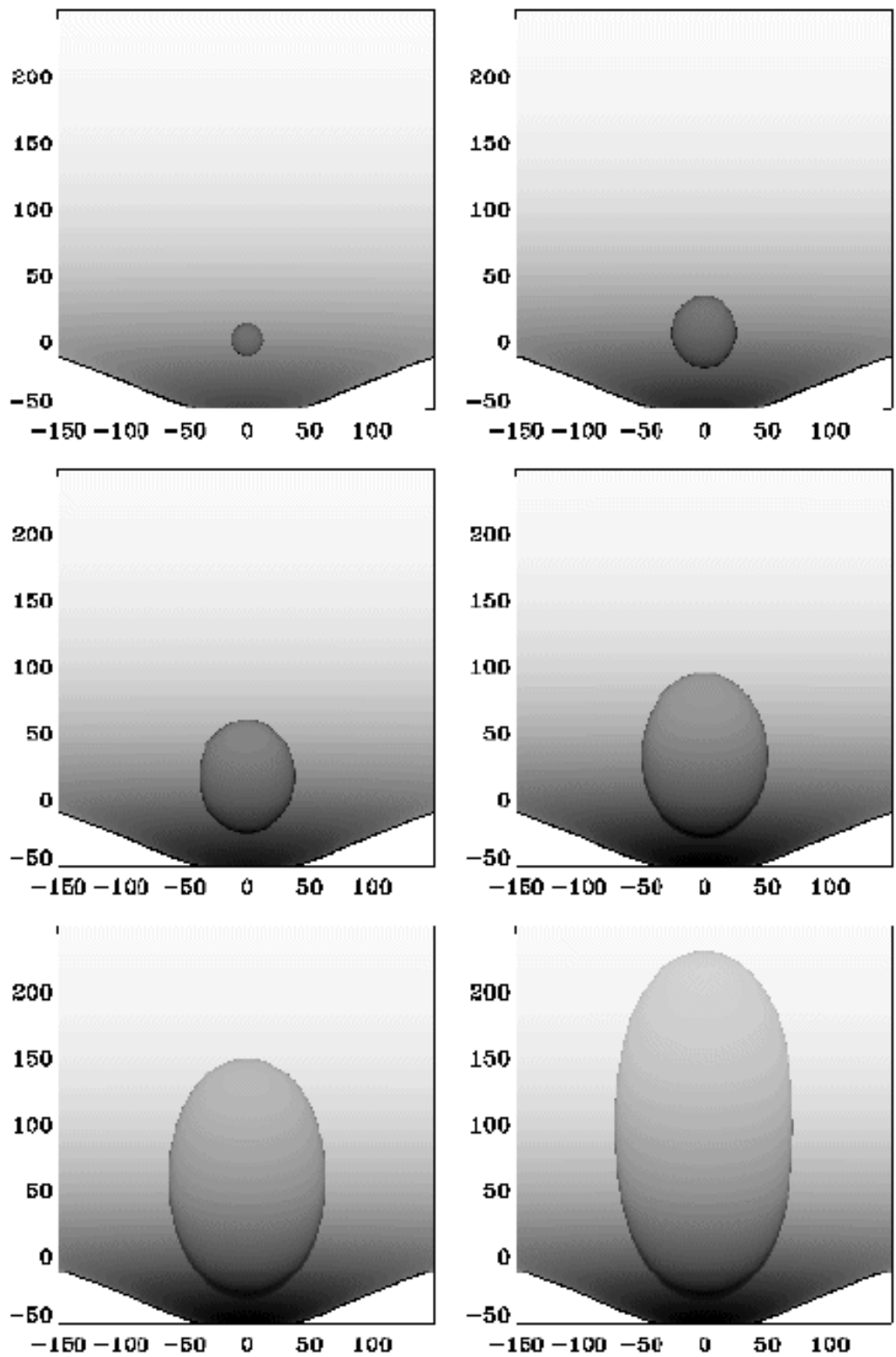}
\caption{(a). Caption on p.33.}
\end{figure}

\clearpage
\addtocounter{figure}{-1}

\begin{figure}
\vspace{9in}
\includegraphics{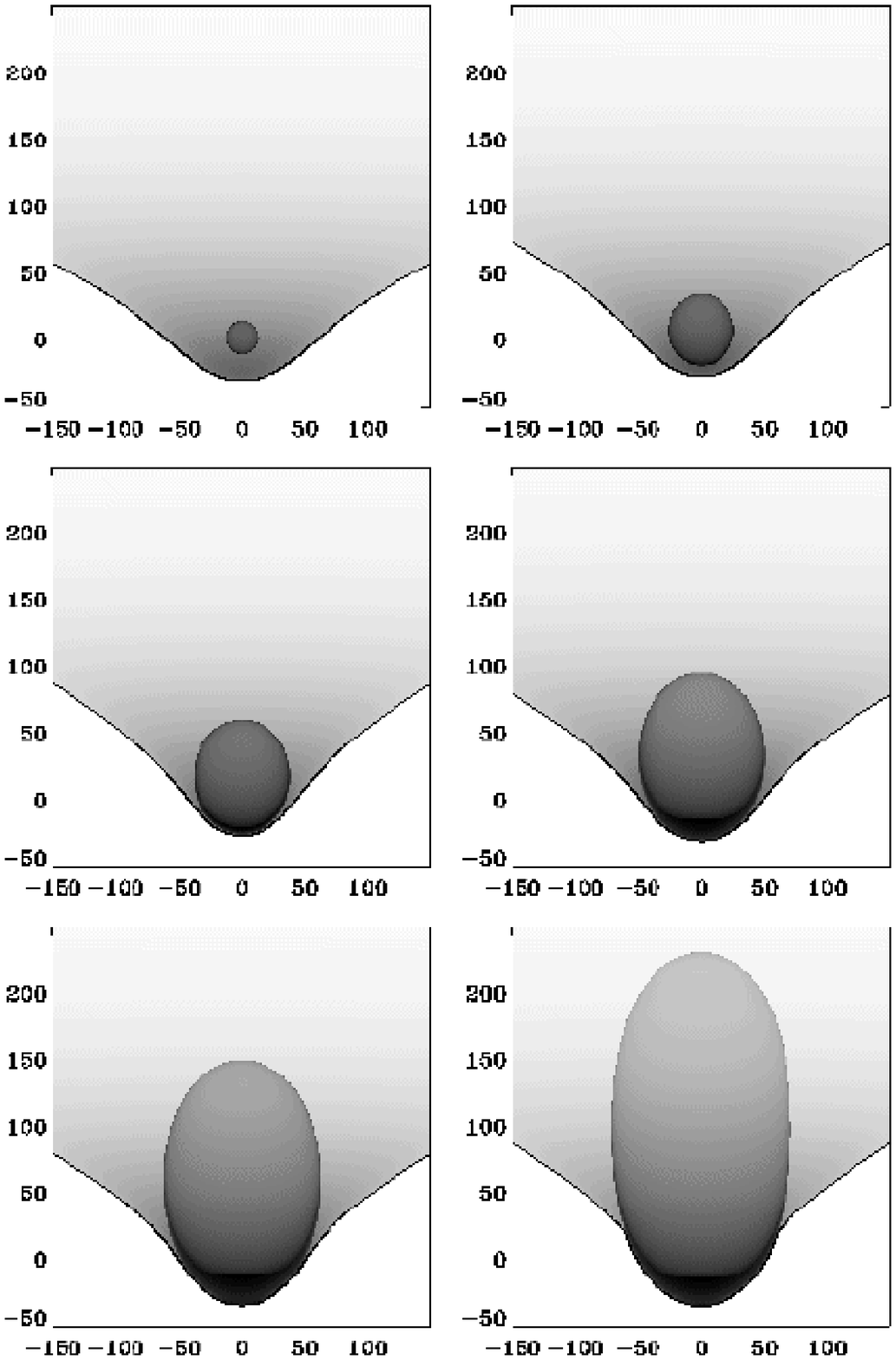}
\caption{(b). Caption on p.33.}
\end{figure}

\clearpage
\addtocounter{figure}{-1}

\begin{figure}
\vspace{9in}
\includegraphics{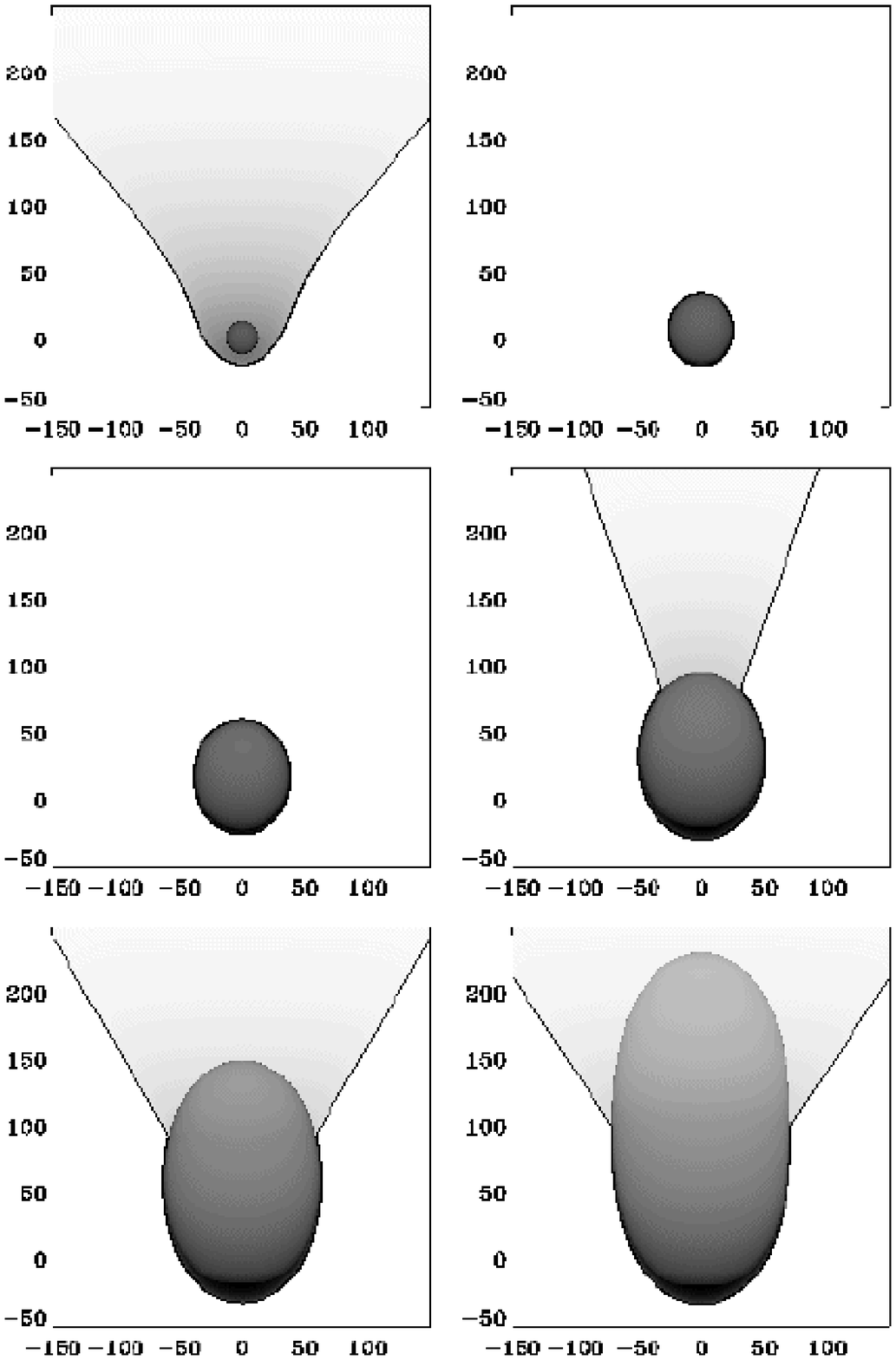}
\caption{(c). Caption on p.33.}
\end{figure}

\clearpage
\addtocounter{figure}{-1}

\begin{figure}
\caption{Relative emission measure across an expanding
bubble ionized by an internal cluster of stars with properties outlined in the
text. Within an exponential atmosphere ($H = 25\,$pc) six stages in evolution
are represented ($\yt$ = 0.5, 1.0, 1.4, 1.7, 1.9, 1.98).
Distances are labeled in pc on both axes,
with the star cluster located at (0,0).
(a) $n_0 = 1\,{\rm cm}^{-3}$.  Note that the ionization front
is always well separated from the bubble.
(b) $n_0 = 5\,{\rm cm}^{-3}$. In this case the increasing column density at the
base of the bubble confines the ionization front at late stages.
(c) $n_0 = 10\,{\rm cm}^{-3}$, representing the best case
scenario for the W4 superbubble.
Note that the ionization front is initially
open towards the top, is confined by the high density swept-up shell at
intermediate epochs, and then breaks out when the bubble gets elongated.
}
\end{figure}

\begin{figure}
\vspace{5in}
\includegraphics{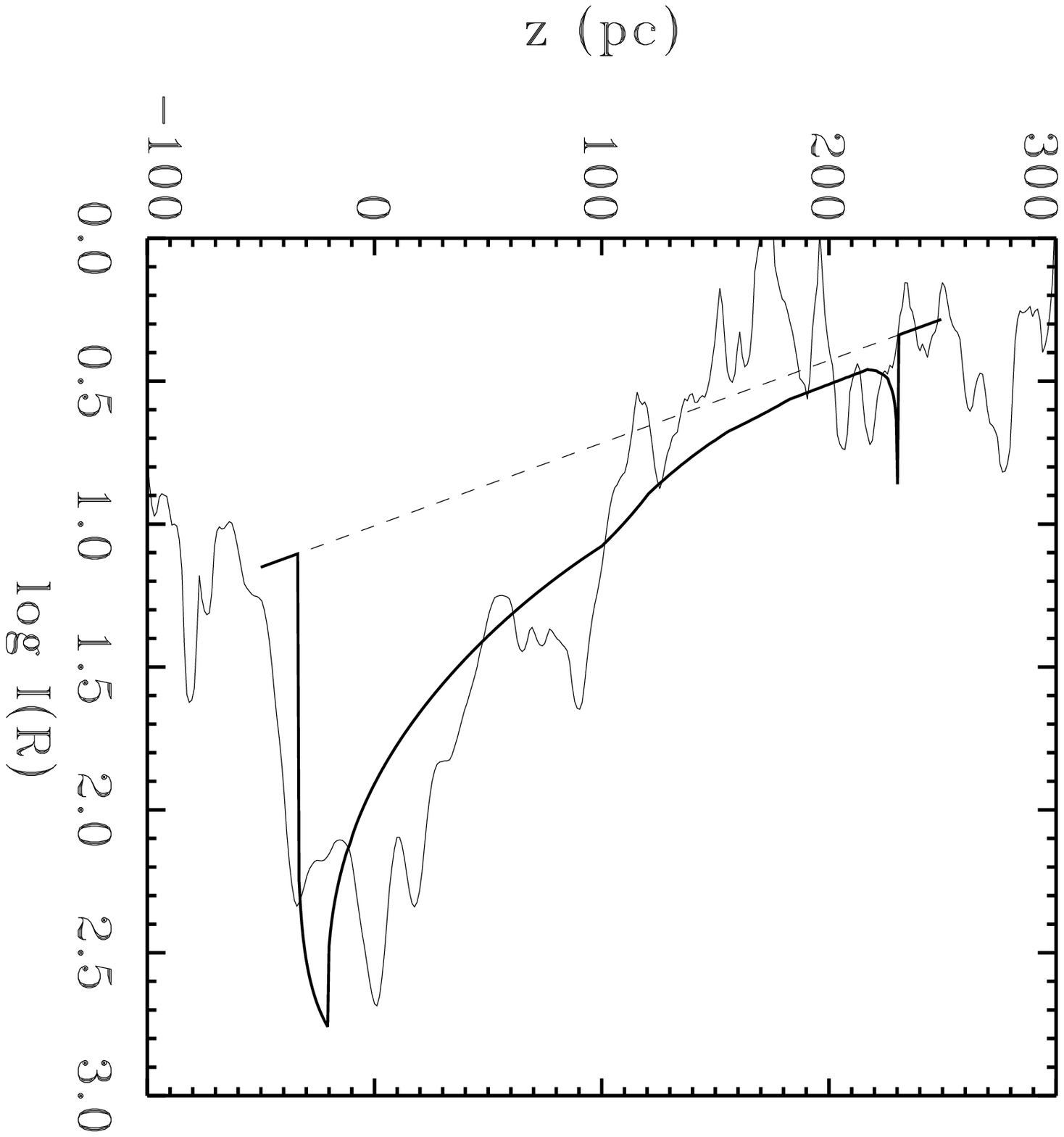}
\caption{
H$\alpha$ intensity profile along the symmetry axis ($r = 0$)
of the superbubble.
Thick line: model described
in Figure 7(c) after conversion to rayleighs, addition of background
flux (dashed line) as described in text, and extinction by 2.1 mag.
Thin line: as observed by DTS along vertical cut centered on the cluster
OCl 352.
}
\end{figure}

\clearpage

\begin{figure}
\vspace{5in}
\includegraphics{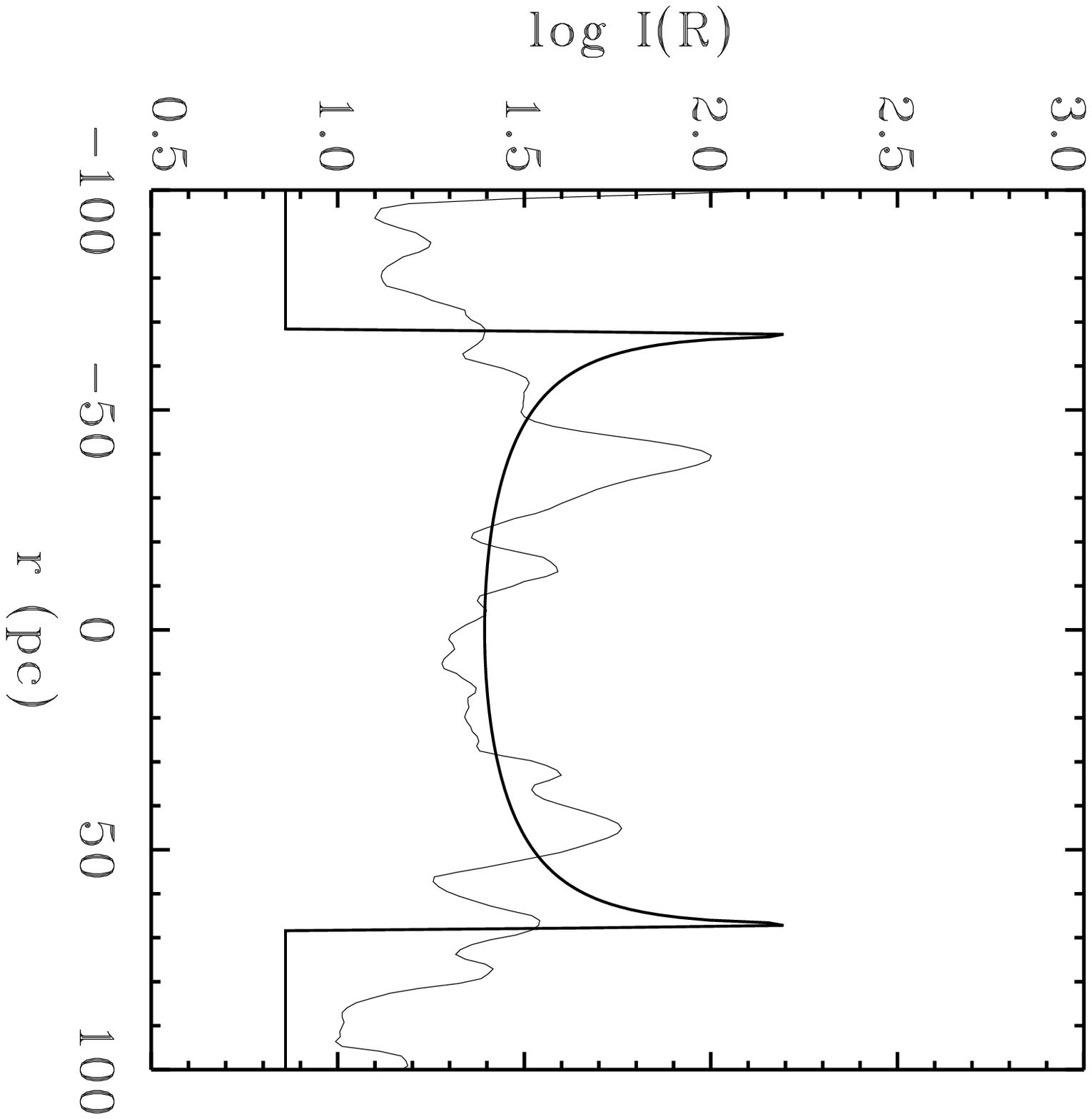}
\caption{
H$\alpha$ intensity profile across the model superbubble
at a height $z = 50\,$pc above the cluster.
Thick line: model described in Figure 7(c)
after conversion to rayleighs, addition of background flux (see text),
and extinction by 2.1 mag.  Thin line: as observed by DTS.
}
\end{figure}

\clearpage

\begin{figure}
\vspace{9in}
\includegraphics{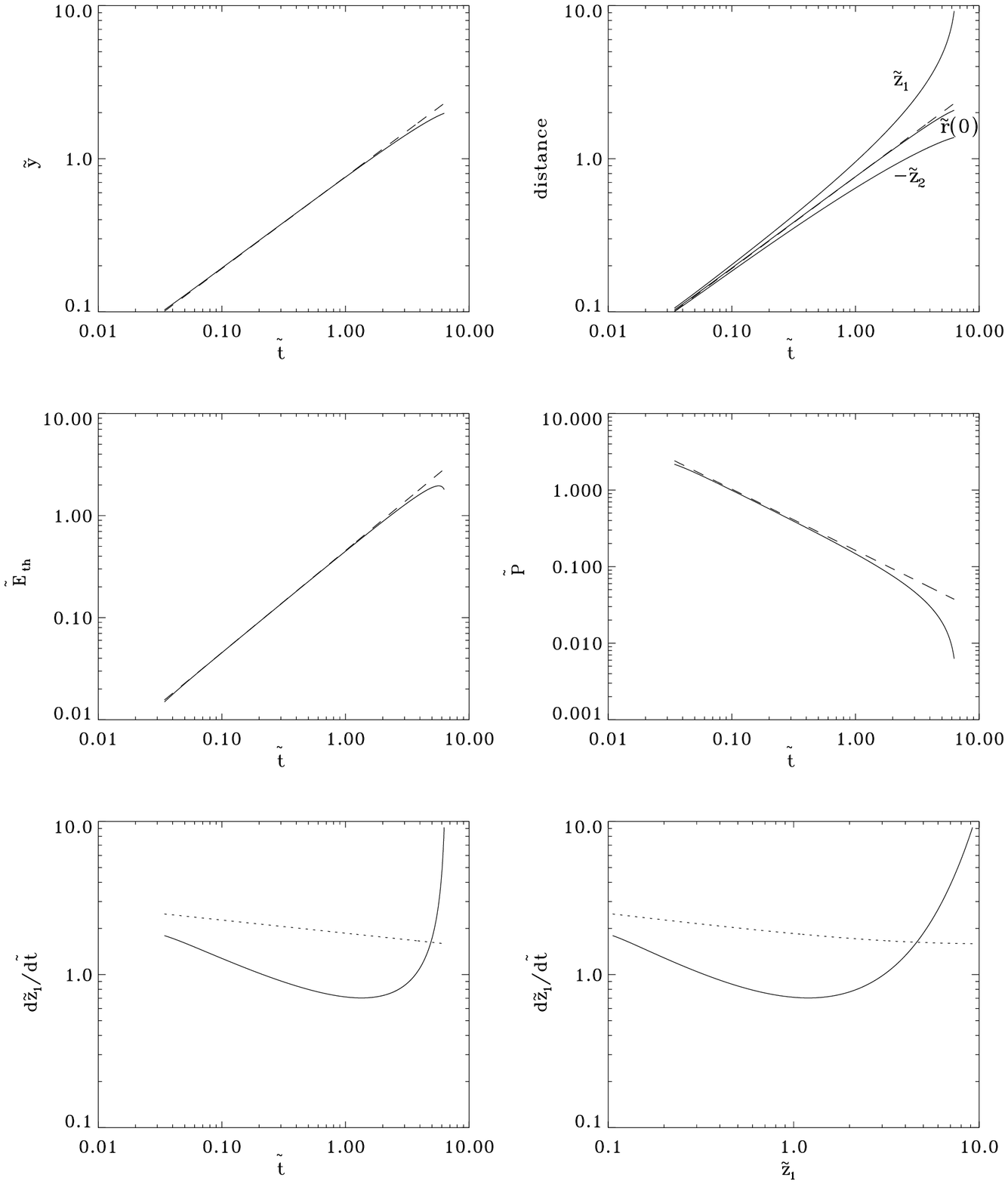}
\caption{Caption on next page.}
\end{figure}

\clearpage
\addtocounter{figure}{-1}

\begin{figure}
\caption{Time evolution of various dimensionless variables in the
wind-blown Kompaneets solution (variables in (a) - (e) plotted
versus the dimensionless time $\ttil$). (a) $\yt$.
For comparison, the dimensionless solution for the radius of a spherical
bubble in a uniform atmosphere, $\tilde{\Rs}$, is shown with a dashed line.
(b) Distance from the wind source to the shock front along various lines of
sight: to the top of the bubble, $\tilde{z}_1$; along
$z=0$, $\tilde{r}(0)$; and to the bottom of the bubble,
$-\tilde{z}_2$. Again, $\tilde{\Rs}$, is shown with a dashed line.
Together, (a) and (b) reveal the important feature $y \simeq r(0) \simeq \Rs$.
(c) Thermal energy $\tilde{E}_{\rm th}$ within the bubble. Dashed line
shows the time-evolution of $\tilde{E}_{\rm th}$ in the spherical solution for a
uniform atmosphere. (d) Pressure $\tilde{P}$ within the bubble. Dashed line
shows the solution for a spherical bubble. (e) Velocity of the top
of the bubble, $d\tilde{z}_1/d\ttil$. The dotted line shows the internal
sound speed $\tilde{c}_{\rm s,i}$
for the parameters $n_0 = 1$ cm$^{-3}$, $L_0 = 3 \times 10^{37}$
ergs s$^{-1}$, and $H = 25$ pc. The Kompaneets approximation is expected to
break down when $d\tilde{z}_1/d\ttil > \tilde{c}_{\rm s,i}$ ($\ttil > 4.9$).
(f) $d\tilde{z}_1/d\ttil$ plotted versus $\tilde{z}_1$.
Dotted line as in (e).
}
\end{figure}

\newpage
\begin{deluxetable}{lc}
\tablewidth{6in}
\tablecaption{UNITS OF PHYSICAL QUANTITIES}

\tablehead{
\colhead{Physical Quantity} & \colhead{Unit} }

\startdata
Length &  $H$ \nl
Time &  $(\rho_0 H^5/L_0)^{1/3}$ \nl
Velocity &   $(L_0/\rho_0 H^2)^{1/3}$ \nl
Acceleration & $(L_0^2/\rho_0^2 H^7)^{1/3}$  \nl
Mass & $\rho_0 H^3$ \nl
Energy & $(\rho_0 L_0^2 H^5)^{1/3}$ \nl
Pressure & $(\rho_0 L_0^2/H^4)^{1/3}$ \nl
\enddata

\end{deluxetable}
\end{document}